\documentclass[a4paper,11pt]{article}
\pdfoutput=1
\usepackage{jcappub} 

\usepackage[subpreambles=true]{standalone}
\usepackage{import}
\usepackage{fullpage}
\usepackage{amsmath,slashed,amssymb,cancel}
\usepackage{graphicx,float,wrapfig,lipsum,subcaption,caption,sidecap}
\usepackage{enumerate,array,multirow,bm,soul}
\usepackage[toc,page]{appendix}
\usepackage[dvipsnames]{xcolor}

\newcommand{\dt}{\,dt}

\newcommand{\tev}{\text{ TeV}}
\newcommand{\gev}{\text{ GeV}}
\newcommand{\mev}{\text{ MeV}}
\newcommand{\kev}{\text{ keV}}
\newcommand{\ev}{\text{ eV}}
\newcommand{\mpl}{M_\text{Pl}}
\newcommand{\fn}[2]{\mathinner{#1\mathopen{\left(#2\right)}}}

\newcommand{\red}[1]{\color{red} #1 \color{black}}
\newcommand{\blue}[1]{\color{blue} #1 \color{black}}
\newcommand{\green}[1]{\color{Green} #1 \color{black}}
\newcommand{\orange}[1]{\color{orange}#1\color{black}}
\newcommand{\purple}[1]{{\color{Purple}#1}}

\title{Constraining Affleck-Dine Leptogenesis after Thermal Inflation}

\author{Seolhwa Kim}
\author{and Ewan D. Stewart}
\affiliation{Department of Physics, KAIST\\ 291 Daehak-ro, Yuseong-gu, Daejeon, Korea}
\emailAdd{liberte@kaist.ac.kr}
\emailAdd{professor@profstewart.org}

\abstract{
		Affleck-Dine leptogenesis after thermal inflation along the $LH_u$ direction requires $m_L^2+m_{H_u}^2 < 0$ up to the AD scale ($|L|\simeq|H_u|\sim 10^9\gev$). We renormalised this condition from the AD scale to the soft supersymmetry breaking scale by solving the renormalisation group equations perturbatively in the Yukawa couplings to obtain a semi-analytic constraint on the soft supersymmetry breaking parameters. We also used a fully numerical method to renormalise the baryogenesis condition and constrained the Minimal Supersymmetric Cosmological Model using the resulting baryogenesis condition and other constraints, specifically, electroweak symmetry breaking, the observed Higgs mass, and the axino dark matter abundance.
}

\begin{document}
\maketitle
\flushbottom

\section{Introduction}\label{s1}

\subsection{Cosmological moduli problem}
\noindent 
Cosmological moduli fields \cite{Coughlan:1983ci,Banks:1993en,deCarlos:1993wie} are scalar fields with Planckian vacuum expectation values. They have a vanishing potential when supersymmetry is unbroken and develop a potential only after supersymmetry breaking. In the early Universe, the moduli fields acquire a potential 
	\begin{equation}\label{moduli potential}
	V_\text{cos}(\Phi) = \rho \fn{f}{\frac{\Phi}{\mpl}} = \frac{\alpha}{2}H^2(\Phi-\Phi_1)^2+ \cdots
	\end{equation}
where $H$ is the Hubble parameter and $\alpha \sim \fn{\mathcal{O}}{1}$. 
 When $H \sim m_\Phi$, their vacuum potential
	\begin{equation}
	V_\text{vac}(\Phi) = m_\text{s}^2\mpl^2 \fn{g}{\frac{\Phi}{\mpl}} = \frac{1}{2}m_\Phi^2\left(\Phi - \Phi_2 \right)^2 + \cdots
	\end{equation}
becomes significant. The minimum is then shifted toward $\Phi_2$, and the moduli fields undergo homogeneous oscillations around $\Phi_2$ with an amplitude $\Phi_0\sim\Phi_1-\Phi_2\sim\mpl$. This coherent oscillation may get critically damped initially as $H \lesssim m_\Phi$, and its large relic density
	\begin{equation} \label{moduli abundance}
	\frac{n_\Phi}{s} \sim \frac{\Phi_0^2}{g_*^{1/4}m_\Phi^{1/2} \mpl^{3/2}}\sim 10^7 \left(\frac{100}{g_*} \right)^\frac{1}{4} \left(\frac{\tev}{m_\Phi} \right)^\frac{1}{2}\left(\frac{\Phi_0}{\mpl} \right)^2
	\end{equation}
where $g_*$ is a relativistic degrees of freedom, leads to a matter domination until its late time decay given by the lifetime
	\begin{gather}
	\tau = \Gamma^{-1} = \left(\frac{\alpha_\Phi m_\Phi^3}{\mpl^2}\right)^{-1} \sim 10^4\text{ s} \,\, \alpha_\Phi^{-1}\left(\frac{\tev}{m_\Phi} \right)^3 
	\end{gather}
and characterized by the low decay temperature 
	\begin{equation}
	T \sim g_*^{-\frac{1}{4}}\Gamma^\frac{1}{2} \mpl^\frac{1}{2} =  \frac{\alpha_\Phi^{1/2}m_\Phi^{3/2}}{g_*^{1/4}\mpl^{1/2}} \sim 10 \kev \,\, \alpha_\Phi^{\frac{1}{2}}\left(\frac{100}{g_*} \right)^\frac{1}{4}\left(\frac{m_\Phi}{\tev} \right)^{\frac{3}{2}}\label{TR}
	\end{equation}
This late time decay into energetic photons or hadronic showers around or after Big Bang nucleosynthesis (BBN) can destroy previously formed nuclei. 
In order not to upset BBN, the decay temperature must satisfy $T \gtrsim 4 \mev$ or the abundance should be low enough \cite{Ellis:1990nb,Kawasaki:2017bqm,Forestell:2018txr} 
	\begin{equation}\label{moduli abundance req}
	\frac{n_\Phi}{s} \lesssim 10^{-12}\left(\frac{\tev}{m_\Phi} \right)
	\end{equation}

There have been some attempts to address this problem. First, if the moduli mass is larger than $10$ to $100\tev$, it decays earlier than BBN and hence does not affect BBN \cite{Acharya:2009zt,Kane:2015jia}. Second, Ref.~\cite{Randall:1994fr} suggested diluting the moduli abundance through a rolling inflation. However, the scale is typically too high so that moduli are regenerated after the inflation. Lastly, thermal inflation \cite{Lyth:1995hj,Lyth:1995ka,YAMAMOTO1986341,YAMAMOTO1985289,ENQVIST1986343,BERTOLAMI1987163,ELLIS1987415,ELLIS1989313} provides enough inflation to dilute moduli at a scale sufficiently low to avoid their regeneration.

\subsection{Thermal inflation}
The low energy effective potential of the flaton is
\begin{equation}
V = V_0 - m_\phi^2|\phi|^2 + \cdots
\end{equation}
where the vacuum expectation value of the flaton satisfies $m_\phi \ll \phi_0 \ll \mpl$, and $V_0 \sim m_\phi^2\phi_0^2 \ll m_\phi^2\mpl^2$ so that there is no slow roll inflation. Thermal inflation starts when the flaton is held at the origin and the thermal energy density ($\sim T^4$) falls below $V_0$. It ends at the critical temperature when the flaton rolls away from the origin ($T\sim m_\phi$). Thus, thermal inflation occurs when the temperature satisfies
\begin{equation}
m_\phi \lesssim T \lesssim V_0^\frac{1}{4}
\end{equation}
The number of e-folds of thermal inflation is therefore
\begin{equation}
N = \ln\frac{a_f}{a_i} = \ln\frac{T_i}{T_f} \simeq \ln\frac{V_0^{1/4}}{m_\phi} \simeq \frac{1}{2}\ln\frac{\phi_0}{m_\phi}\sim 10 
\end{equation}
for $\phi_0 \sim 10^{11}\gev$ and $m_\phi \sim 1\tev$ where $a_f$ and $a_i$ are scale factors at the beginning and end of thermal inflation respectively. This moderately redshifts the density perturbation from the primordial inflation.

After thermal inflation, the flaton decays with $\Gamma = \beta m_\phi^3/\phi_0^2$. Assuming decay products thermalize promptly, the decay temperature is
\begin{gather}\label{Td}
T_d \sim g_*^{-\frac{1}{4}}\Gamma^{\frac{1}{2}}\mpl^{\frac{1}{2}} \simeq 1\gev \,\, \beta^{\frac{1}{2}}\left(\frac{100}{g_*}\right)^{-\frac{1}{4}}\left(\frac{10^{11}\gev}{\phi_0} \right) \left(\frac{m_\phi}{\tev} \right)^{\frac{3}{2}}
\end{gather}
For the decay to precede BBN, one requires $T_d \gtrsim 1 \mev$, corresponding to $\phi_0 \lesssim 10^{14}\gev$ for $m_\phi \sim 1\tev$.

To solve the moduli problelm with thermal inflation, one should consider both pre-existing moduli and the moduli regenerated after thermal inflation. First, the entropy released after thermal inflation dilutes the pre-existing moduli fields by a factor
		\begin{align}\label{entropy increase}
		\Delta & = \frac{s_d R_d^3}{s_c R_c^3} \sim \frac{{\rho_\phi}_d R_d^3/T_d}{{g_*}_cR_c^3T_c^3 } \sim
		\frac{V_0}{{g_*}_cT_c^3T_d}
		\sim 10^{17} \left(\frac{100}{{g_*}_c} \right)\left(\frac{V_0^{1/4}}{10^7\gev} \right)^4 \left(\frac{\tev}{T_c} \right)^3 \left(\frac{\gev}{T_d}\right)
		\end{align}
where the subscript $b$, $c$ and $d$ denote at the beginning of and the end of thermal inflation, and at the decay of the flaton respectively. Combined with Eq.~\eqref{moduli abundance}, the moduli abundance is diluted to
		\begin{equation}
		\left.\frac{n_\Phi}{s}\right|_d = \frac{1}{\Delta}\left.\frac{n_\Phi}{s}\right|_a \sim 10^{-10} \left(\frac{g_*}{100} \right)^\frac{3}{4}\left(\frac{\tev}{m_\Phi} \right)^\frac{1}{2} \left(\frac{\Phi_0}{\mpl}\right)^2 \left(\frac{10^7\gev}{V_0^{1/4}} \right)^4 \left(\frac{T_c}{\tev} \right)^3 \left(\frac{T_d}{\gev} \right).
		\end{equation}
where the subscript $a$ denotes when $H\sim m_\Phi$. Since this is greater than the bound in Eq.~\eqref{moduli abundance req}, a sufficient dilution may require double thermal inflation \cite{Lyth:1995ka,Felder:2007iz,Kim:2008yu,Choi:2012ye}.
		
Meanwhile, the thermal inflationary potential energy is moduli field-dependent. This appears in the potential of the moduli fields as
		\begin{align}
		V &= \frac{1}{2}m_\Phi^2(\Phi - \Phi_2)^2 - \frac{\alpha'V_0}{\mpl}\Phi +\cdots
		\end{align}
Hence, the moduli fields begin to oscillate after thermal inflation with an amplitude
		\begin{equation}\label{moduli shift}
		\delta\Phi \sim \frac{\alpha' V_0}{m_\Phi^2\mpl}
		\end{equation}
However, the corresponding moduli abundance
		\begin{align}
		\frac{n_\Phi}{s} &\sim \frac{{n_\Phi}_c}{s_d}\frac{{n_\Phi}_d}{{n_\Phi}_c} \sim \frac{m_\Phi\delta\Phi^2}{s_d}\frac{\rho_d}{V_0} \sim \frac{\alpha'^2V_0 T_d}{m_\Phi^3\mpl^2}\\
		& = 10^{-16} \,\,\alpha'^2\left(\frac{V_0^{1/4}}{10^7\gev} \right)^4 \left(\frac{T_d}{\gev} \right)  \left(\frac{\tev}{m_\Phi} \right)^3
		\end{align}
is safe.


\subsection{Affleck-Dine leptogenesis after thermal inflation}
The low scale of thermal inflation $V^{1/4}_0 \sim 10^7 \gev$ and the low reheat temperature $T_d \sim  1\gev$ make it hard to realize baryogenesis. For example, GUT baryogenesis and right-handed neutrino leptogenesis rely on particles with mass $M_\text{GUT} \sim 10^{16}\gev$ and $m_{\bar{\nu}}\gtrsim 10^9\gev$ \cite{Davidson:2002qv}, which decay before thermal inflation. Hence, any baryon asymmetries from their decays would be diluted by thermal inflation to a negligible amount. Likewise, the standard Affleck-Dine (AD) mechanism \cite{Affleck:1984fy,Dine:2003ax} in gravity-mediated supersymmetry breaking scenarios\footnote{We refer readers to Ref.~\cite{Kasuya:2001tp} for the AD baryogenesis assuming gauge mediated supersymmetry breaking.} occurs before thermal inflation when the Hubble parameter is comparable to the sparticle masses. These difficulties suggest that baryogenesis should occur during the thermal inflationary era and early works in this direction are given in Ref.~\cite{Lazarides:1985ja,Yamamoto:1986jw,Mohapatra:1986dg}.

This paper focuses on AD leptogenesis after thermal inflation \cite{Stewart:1996ai,Jeong:2004hy,Kawasaki:2006py,Felder:2007iz,Kim:2008yu,Choi:2009qd,Park:2010qd} implemented in the Minimal Supersymmetric Cosmological Model (MSCM) \cite{Kim:2008yu}. The MSCM is a minimal implementation of supersymmetry, thermal inflation and baryogenesis with the QCD axion \cite{Kawasaki:2014sqa,Klaer:2017ond} and axino \cite{Covi:2009pq,Choi:2013lwa} as dark matter. The superpotential of the MSCM is
	\begin{equation}\label{mscm}
	W = \lambda_u QH_u\bar{u} + \lambda_d QH_d\bar{d} + \lambda_e LH_d\bar{e} + \frac{1}{2}\kappa_\nu\left (LH_u\right)^2 + \kappa_\mu\phi^2 H_uH_d + \lambda_\chi\phi\chi\bar{\chi} 
	\end{equation}
where $\frac{1}{2}\kappa_\nu\left(LH_u\right)^2$ provides the neutrino mass
		\begin{equation}
		m_\nu = |\kappa_\nu H_u^2| = |\kappa_\nu| v^2\sin^2\beta \label{rhn mass}
		\end{equation} 
and $\mu=\kappa_\mu\phi_0^2$ is the MSSM $\mu$-parameter. The coupling $\lambda_\chi$ renormalises the flaton's mass to become negative at the origin and couples the flaton to the thermal bath, holding it at the origin during thermal inflation. As well as being the flaton whose potential drives thermal inflation, $\phi$ also acts as the Peccei-Quinn field containing the QCD axion.

In the MSCM, we obtain a lepton asymmetry by the AD mechanism along the $LH_u$ direction. For the $LH_u$ direction to have an initially large field value, it is necessary to have
		\begin{equation}\label{pre-bg}
		m^2_{LH_u} \equiv \frac{1}{2}\left(m^2_{L} + m^2_{H_u}\right) < 0
		\end{equation}
for $|L|\sim|H_u| \lesssim \text{AD}$ scale for some lepton generation. Eq.~\eqref{pre-bg} is most easily satisfied for the third generation and so we take $L=L_3$ in this paper. Moreover, since $m_{L_3H_u}^2$ becomes more negative at lower scales, it is sufficient to require
		\begin{equation}
		\left.m^2_{L_3H_u}\right|_\text{AD} < 0 \label{bg condition}
		\end{equation}
We call Eq.~\eqref{bg condition} the baryogenesis condition. 

With the ansatz
		\begin{equation}
		L_3 = \begin{pmatrix}
		l \\ 0
		\end{pmatrix}, H_u = \begin{pmatrix}
		0 \\ h_u 
		\end{pmatrix}, H_d = \begin{pmatrix}
		h_d \\ 0 
		\end{pmatrix} , \phi = \phi
		\end{equation}\label{parametrization}
and other fields zero, the MSCM potential reduces to	
		\begin{align}
		\nonumber V & = V_0  + \tilde{m}_\phi^2|\phi|^2 + m_L^2|l|^2 + m_{H_u}^2|h_u|^2 + m_{H_d}^2|h_d|^2 +  \left(\frac{1}{2}A_\nu\kappa_\nu l^2h_u^2 - B \kappa_\mu\phi^2h_uh_d + \text{c.c.}\right)\\
		&\quad + \left|\kappa_\nu lh_u^2\right|^2 + \left|\kappa_\nu l^2h_u - \kappa_\mu\phi^2h_d\right|^2 + \left|\kappa_\mu\phi^2h_u\right|^2 + \left|2\kappa_\mu\phi h_uh_d\right|^2 +\frac{g^2}{2}\left (|h_u|^2 - |h_d|^2 - |l|^2\right)^2
		\end{align}
where $\tilde{m}_\phi^2 (\phi)$ is the renormalised mass of the flaton which runs from $\tilde{m}_\phi^2>0$ at large field values to $\tilde{m}_\phi^2 < 0 $ near the origin.
We assume all fields are initially held at the origin due to finite temperature effects. While the flaton is held at the origin during thermal inflation, the temperature drops and the $LH_u$ flat direction becomes unstable due to Eq.~\eqref{bg condition}. The corresponding minimum
			\begin{gather}
			A_\nu\kappa_\nu l^2h_u^2 = -|A_\nu\kappa_\nu l^2h_u^2|\\
			|l| \simeq |h_u| \simeq \sqrt{\frac{|A_\nu|+\sqrt{|A_\nu|^2-12m_{LH_u}^2}}{6|\kappa_\nu|}} \simeq 10^9\gev \sqrt{\left(\frac{|m_{LH_u}|}{\tev}\right) \left(\frac{10^{-2}\ev}{m_\nu} \right)}
			\end{gather}

provides the initial condition for Affleck-Dine leptogenesis.

After thermal inflation ends, as the flaton begins to roll away from the origin, two terms in the potential, $B\kappa_\mu\phi^2h_uh_d$ and $(\kappa_\nu l^2h_u)^{*}\kappa_\mu\phi^2 h_d +\text{c.c.}$, give rise to a non-zero field value of $h_d$. When $\phi$ approaches to $\phi_0$, the term $\left|\kappa_\mu\phi^2h_u\right|^2$ gives a positive contribution to the mass squared in the $LH_u$ direction at the origin, pulling $l, h_u$ and $h_d$ towards the origin.
Meanwhile, the terms $B\kappa_\mu\phi^2h_uh_d$ and $(\kappa_\nu l^2h_u)^{*}\kappa_\mu\phi^2 h_d +\text{c.c.}$ tilt the potential in the phase direction. This changes the phase of $lh_u$ and hence produces a lepton asymmetry.
		
The amplitude of the homogeneous mode is damped due to preheating and friction induced by the thermal bath. Therefore, the lepton number violating terms become less significant so that the lepton number is conserved. After the AD field's preheating and decay, the associated partial reheat temperature allows sphaleron processes that convert the lepton number to a baryon number.

\subsection{This paper}
\noindent
In Section \ref{s2}, we will solve the renormalisation group equations to translate the baryogenesis condition from the AD scale to the soft supersymmetry breaking scale and obtain semi-analytic constraints on the soft supersymmetry breaking parameters. Then, we will compare the semi-analytic formula to results using the numerical package FlexibleSUSY \cite{Athron:2017fvs}. In Section \ref{s3}, we will assume CMSSM boundary conditions and combine the baryogenesis constraint with other constraints--electroweak symmetry breaking, the Higgs mass, the axino cold dark matter abundance and the stability of the $H_uH_d$ direction.

In Appendix \ref{rgimprovement}, we will address the connection between a field value and the renormalisation scale through the renormalisation group improvement. In Appendix \ref{perturbation}, we give the detailed calculation for the semi-analytic formula of the baryogenesis condition.

\section{Renormalisation of the baryogenesis condition}\label{s2}

The baryogenesis condition
\begin{equation}
\left.m^2_{L_3H_u}\right|_\text{AD} < 0 \tag{\ref{bg condition}}
\end{equation}
applies to the large field value, $|L_3|\sim|H_u|\sim 10^9 \gev$. 
In Appendix~\ref{rgimprovement}, we will explain how the renormalisation group improvement connects the renormalisation of couplings in field space with the renormalisation with respect to the renormalisation scale. Resorting to this, we solve the renormalisation group equations with respect to the renormalisation scale and obtain the baryogenesis condition imposed at the AD field value expressed in terms of the parameters at the soft supersymmetry breaking scale, or alternatively, in terms of the universal CMSSM GUT parameters.

The relevant renormalisation group equations are

\begin{equation}
x = \frac{1}{8\pi^2}\log\frac{\mu}{m_s}
\end{equation}
\begin{eqnarray} \label{RGEg1}
\frac{d}{dx} \frac{1}{g_1^2} & = & - \frac{33}{5}
\\ \label{RGEg2}
\frac{d}{dx} \frac{1}{g_2^2} & = & - 1
\\ \label{RGEg3}
\frac{d}{dx} \frac{1}{g_3^2} & = & 3
\end{eqnarray}
\begin{equation}
M_i \propto g_i, \quad i=1,2,3
\end{equation}
\begin{eqnarray} \label{RGElt}
\frac{d}{dx} \frac{1}{\lambda_t^2} & = & \left( \frac{16}{3} g_3^2 + 3 g_2^2 + \frac{13}{15} g_1^2 \red{-\lambda_b^2} \purple{-\lambda_\nu^2} \right) \frac{1}{\lambda_t^2} - 6
\\ \label{RGElb}
\frac{d}{dx} \frac{1}{\lambda_b^2} & = & \left( \frac{16}{3} g_3^2 + 3 g_2^2 + \frac{7}{15} g_1^2 - \lambda_t^2 \orange{-\lambda_\tau^2} \right) \frac{1}{\lambda_b^2} - 6
\\ \label{RGEltau}
\frac{d}{dx} \frac{1}{\lambda_\tau^2} & = & \left( 3 g_2^2 + \frac{9}{5} g_1^2 - 3 \lambda_b^2 \purple{-\lambda_\nu^2} \right) \frac{1}{\lambda_\tau^2} - 4
\end{eqnarray}
\begin{eqnarray} \label{RGEAt}
\frac{d}{dx} A_t & = & 6 \lambda_t^2 A_t \red{+ \lambda_b^2 A_b} \purple{+\lambda_\nu^2A_\nu} + \frac{16}{3} g_3^2 M_3 + 3 g_2^2 M_2 + \frac{13}{15} g_1^2 M_1
\\ \label{RGEAb}
\frac{d}{dx} A_b & = & \lambda_t^2 A_t + 6 \lambda_b^2 A_b \orange{+ \lambda_\tau^2 A_\tau} + \frac{16}{3} g_3^2 M_3 + 3 g_2^2 M_2 + \frac{7}{15} g_1^2 M_1
\\
\frac{d}{dx} A_\tau & = & 3 \lambda_b^2 A_b + 4 \lambda_\tau^2 A_\tau \purple{+\lambda_\nu^2A_\nu} + 3 g_2^2 M_2 + \frac{9}{5} g_1^2 M_1 \label{RGEAtau}
\end{eqnarray}

\begin{eqnarray} \label{RGEmlt}
\frac{d}{dx} D_t
& = & 6 \lambda_t^2 A_t^2 \red{+ \lambda_b^2 A_b^2} \purple{+\lambda_\nu^2A_\nu^2}
- \frac{32}{3} g_3^2 M_3^2 - 6 g_2^2 M_2^2 - \frac{26}{15} g_1^2 M_1^2 
+ 6 \lambda_t^2 D_t
\red{+ \lambda_b^2 D_b} \purple{+ \lambda_\nu^2D_\nu^2}
\\ \label{RGEmlb}
\frac{d}{dx} D_b & = & \lambda_t^2 A_t^2 + 6 \lambda_b^2 A_b^2  \orange{+ \lambda_\tau^2 A_\tau^2}
- \frac{32}{3} g_3^2 M_3^2 - 6 g_2^2 M_2^2 - \frac{14}{15} g_1^2 M_1^2 + \lambda_t^2 D_t + 6 \lambda_b^2 D_b \orange{+ \lambda_\tau^2 D_\tau}
\\ \label{RGEmltau}
\frac{d}{dx} D_\tau & = & 3 \lambda_b^2 A_b^2  + 4 \lambda_\tau^2 A_\tau^2 \purple{+\lambda_\nu^2A_\nu^2}
- 6 g_2^2 M_2^2 - \frac{18}{5} g_1^2 M_1^2
+ 3 \lambda_b^2 D_b + 4 \lambda_\tau^2 D_\tau \purple{+\lambda_\nu^2D_\nu}
\end{eqnarray}

\begin{equation}\label{RGEmlhu}
\frac{d}{dx} \left( m_{L_3}^2 + m_{H_u}^2 \right) = 3 \lambda_t^2 A_t^2 \green{+ \lambda_\tau^2 A_\tau^2} + \purple{+2\lambda_\nu^2A_\nu^2}
- 6 g_2^2 M_2^2 - \frac{6}{5} g_1^2 M_1^2
+ 3 \lambda_t^2 D_t \green{+ \lambda_\tau^2 D_\tau} \purple{+2\lambda_\nu^2D_\nu^2}
\end{equation}

where
\begin{gather}
D_t  \equiv m_{H_u}^2 + m_{Q_3}^2 + m_{\bar{t}}^2, \quad 
D_b \equiv m_{H_d}^2 + m_{Q_3}^2 + m_{\bar{b}}^2,\\
D_\tau  \equiv m_{H_d}^2 + m_{L_3}^2 + m_{\bar\tau}^2, \quad
D_\nu \equiv m_{H_u}^2+m_{L_3}^2+m_{\bar\nu}^2
\end{gather}

\noindent
\purple{Purple} terms are from the right-handed neutrino which we neglect until Section \ref{rhn},

These equations take the form of
		\begin{equation}
		\frac{d}{dx}y = f(x)y + g(x) \label{formula1}
		\end{equation}
which has the solution 
		\begin{equation}
		y(x) = e^{\int^x_0 f(t) dt}\int^x_0 g(t)e^{-\int^t_0f(t')dt'} dt \label{formula2}
		\end{equation}
Using this, the gauge couplings and the gaugino masses can be solved exactly
		\begin{eqnarray}
		g_1^2 &=& \frac{g_1^2(0)}{1-\frac{33}{5}\fn{g_1^2}{0}x} \label{Solg1}, \quad g_2^2 = \frac{g_2^2(0)}{1-\fn{g_2}{0}x}, \quad g_3^2 = \frac{g_3^2(0)}{1+3\fn{g_3^2}{0}x}\\
		M_1 &=& \frac{M_1(0)}{1-\frac{33}{5}\fn{g_1^2}{0}x},\quad M_2 = \frac{M_2(0)}{1-\fn{g_2^2}{0}x}, \quad M_3 = \frac{M_3(0)}{1+3\fn{g_3^2}{0}x} \label{SolM3}
		\end{eqnarray}
\noindent		
Noting the hierarchy in Yukawa couplings
		\begin{equation}\label{yukawa}
		\begin{split}
		\lambda_t^2 \simeq 1.0 \times \left(1+\frac{1}{\tan^2\beta} \right), \quad \lambda_b^2 \simeq 5.6\times 10^{-4}(1+\tan^2\beta), \quad \lambda_\tau^2 \simeq 1.0\times 10^{-4}(1+\tan^2\beta)
		\end{split}
		\end{equation}
i.e. $ \lambda_\tau < \lambda_b < \lambda_t \sim 1.0 $ for $\tan\beta < 42$, we will use perturbation in $\lambda_b$ and $\lambda_\tau$ to solve the remaining renormalisation group equations for $m_{L_3H_u}^2$.

\subsection{Zeroth order semi-analytic result}
\begin{table}[htb!]
	\centering
	\begin{tabular}{c c c c c}
		$\lambda_t^{(0)}$ & $\rightarrow$ &\color{red} $\lambda_b$ &$\rightarrow$ & \color{Green}$\lambda_\tau, \color{blue}\lambda_t^{(1)}$\\
		$\downarrow$ & &$\downarrow$& & $\downarrow$ \\
		$A_t^{(0)}$ & $\rightarrow$ &\color{red} $A_b$ & $\rightarrow$ &\color{Green} $A_\tau, \color{blue}A_t^{(1)}$\\
		$\downarrow$ & &$\downarrow$ & &$\downarrow$\\
		$D_t^{(0)}$ & $\rightarrow$ & \color{red}$D_b$ & $\rightarrow$ & $\color{Green}D_\tau,\color{blue}D_t^{(1)}$\\
		$\downarrow$& & & &$\downarrow$ \\
		$ \left(m_{L_3}^{2}+m_{H_u}^{2}\right)^{(0)}$& & & &$\left(m_{L_3}^{2}+m_{H_u}^{2}\right)^{(1)}$ 
	\end{tabular}
	\caption{Sequence of solving for $m_{L_3H_u}^2$ perturbatively in $\lambda_b, \lambda_\tau$}
	\label{sequence}
\end{table}

\noindent
In the zeroth order, we set $\lambda_b = \lambda_\tau=0$. This amounts to neglecting the colored terms in Eqs.~\eqref{RGElt}, \eqref{RGEAt}, \eqref{RGEmlt} and \eqref{RGEmlhu}. First, we solve Eq.~\eqref{RGElt} for $\lambda_t^{(0)}$. Then, we substitute the resulting $\lambda_t^{(0)}$ into Eq.~\eqref{RGEAt} and solve for $A_t^{(0)}$. Next, we substitute the resulting $A_t^{(0)}$ and $\lambda_t^{(0)}$ into Eq.~\eqref{RGEmlt} and solve for $D_t^{(0)}$. Finally, we substitute $\lambda_t^{(0)},A_t^{(0)}$ and $D_t^{(0)}$ into Eq.~\eqref{RGEmlhu} and solve for $m_{L_3H_u}^{2(0)}$. The sequence above can be summarized as $\lambda_t^{(0)} \rightarrow A_t^{(0)} \rightarrow D_t^{(0)} \rightarrow m^{2(0)}_{LH_u}$ (first column of Table~\ref{sequence}). We put the detailed calculations in Appendix~\ref{a2-1}. The resulting zeroth order analytic expression is
\begin{gather} 
\left.\left(m^2_{L_3} + m^2_{H_u}\right)^{(0)}\right|_\text{AD} = m_{L_3}^2(0) + m_{H_u}^2(0) + \sum_X \alpha_X^{(0)}\fn{X}{0},  \quad \text{with } X \in\{D_t,A_t^2,A_tM_i,M_iM_j \,|\, i=1,2,3 \} \label{expansion0}
\end{gather}
where the coefficients $\alpha_X^{(0)}$ are functions of $g_i(0), \lambda_t(0)$ and $x$ and given in Tables~\ref{tanb5} to \ref{tanb30} in Appendix~\ref{a2-3}. For example, the baryogenesis condition for $\tan\beta = 20$ and the AD scale at $10^9$ GeV ($x = 0.175$) is 
	\begin{align}\label{ex formula0}
	0 > \nonumber \left.\left(m_{L_3}^{2} + m_{H_u}^{2}\right)^{(0)}\right|_\text{AD}  &= m_{L_3}^{2} + m_{H_u}^{2}  
	\nonumber + 0.49 A_t^2+A_t (0.01 M_1+0.08 M_2+0.25 M_3)+0.30 D_t\\
	&\quad {} -0.07 M_1^2-0.44 M_2^2-0.10 M_3^2+0.03 M_2 M_3
	\end{align}
where all parameters on the right-hand side are evaluated at the soft supersymmetry breaking scale.

Since the $M_1$ contribution is small compared to other numerical coefficients in Table~\ref{tanb5} to \ref{tanb30}, we set $M_1=0$ to give the simplified zeroth order semi-analytic expression
	\begin{gather}\label{simplified formula}
\left.\left(m_{L_3}^{2} + m_{H_u}^{2}\right)_\text{S}^{(0)}\right|_\text{AD}  =  m_{L_3}^2(0) + m_{H_u}^2(0) + \sum_X \alpha_X^{(0)}\fn{X}{0}, \quad \text{with } X\in \{D_t, A_t^2, A_tM_i, M_iM_j \,|\, i=2,3 \}  
\end{gather}
For example, for $\tan\beta = 20$ and the AD scale at $10^9\gev$, the simplified zeroth order baryogenesis condition is
\begin{align}
0> \nonumber \left.\left(m_{H_u}^{2} + m_{L_3}^{2}\right)^{(0)}_\text{S}\right|_\text{AD}  &= m_{L_3}^{2}+m_{H_u}^{2}  
+ 0.49 A_t^2+A_t (0.08 M_2+0.25 M_3)+0.30 D_t\\
&\quad -0.44 M_2^2-0.10 M_3^2+0.03 M_2 M_3 \label{simp anal bg}
\end{align}

Alternatively, one can replace $x=0$ in the integral domain of Eq.~\eqref{mlhu0} with the GUT scale to express the zeroth order semi-analytic formula of the baryogenesis condition in terms of the universal CMSSM GUT scale parameters as
\begin{align}
0>\left.\left(m_{L_3}^2+m_{H_u}^2\right)^{(0)}\right|_\text{AD} &= 1.47 m_0^2 - 0.11 A_0^2 +0.13A_0 M_{1/2} +0.30 M_{1/2}^2 \label{gut formula}
\end{align}

\subsection{First order semi-analytic result}
\noindent
At first order, we follow the second and third columns in Table~\ref{sequence}. First, we set $\lambda_\tau =  0$ (neglecting \orange{orange } terms) and solve Eqs.~\eqref{RGElb}, \eqref{RGEAb} and \eqref{RGEmlb} for $\lambda_b\rightarrow A_b\rightarrow D_b$ using the zeroth order values of $\lambda_t^{(0)},A_t^{(0)}$ and $D_t^{(0)}$ in the same way we solved for $\lambda_t^{(0)}\rightarrow A_t^{(0)}\rightarrow D_t^{(0)}$ at zeroth order. Next, we substitute $\lambda_b$ and $\lambda_t^{(0)}$ into Eq.~\eqref{RGEltau} and solve for $\lambda_\tau$. Then, we follow the same sequence in the previous step to solve Eqs.~\eqref{RGEltau}, \eqref{RGEAtau} and \eqref{RGEmltau} for $\lambda_\tau \rightarrow A_\tau \rightarrow D_\tau$ (\green{green} in the third column of Table~\ref{sequence}). In the same way, we solve Eqs.~\eqref{RGElt}, \eqref{RGEAt} and \eqref{RGEmlt} for $\lambda_t^{(1)},A_t^{(1)}$ and $D_t^{(1)}$ with $\lambda_b, A_b,$ and $D_b$ calculated before (\blue{blue} in the third column of Table~\ref{sequence}). Finally, we combine $D_b,D_\tau$ and $D_t^{(1)}$ to compute $m_{L_3H_u}^{2(1)}$.


We put the detailed calculations in Appendix~\ref{a2-2} and the resulting zeroth and first order expressions of $m^2_{L_3} + m^2_{H_u}$ in Tables~\ref{tanb5} to \ref{tanb30}. For example, for $\tan\beta = 20$ and the AD scale at $10^9$ GeV ($x = 0.175$), the first order semi-analytic formula of the baryogenesis condition is
\begin{equation}
	\begin{split}
	0 > \left.\left(m_{L_3}^{2} + m_{H_u}^{2}\right)^{(1)} \right|_\text{AD} &= m_{L_3}^{2} + m_{H_u}^{2} + 0.49 A_t^2 +A_t (0.01 M_1+0.08 M_2+0.26 M_3) +0.31 D_t \nonumber \\
	&\quad -0.07 M_1^2-0.42 M_2^2 - 0.10 M_3^2 +0.02 M_1 M_2 + 0.03 M_2 M_3 + 0.01 M_1 M_3 \nonumber \\
	&\quad +0.01 A_b M_2 +0.01 A_\tau^2 + 0.01D_\tau \label{ex formula1}
	\end{split}
	\end{equation}
From Tables~\ref{tanb5} to \ref{tanb30}, the coefficients of the zeroth and the first order semi-analytic formulae are close, which gives confidence to this perturbative calculation.

\subsection{Comparison with FlexibleSUSY}

\begin{table}[htbp]
	\centering
	\setlength\extrarowheight{5pt}
	\begin{tabular}{c|c|c|c|c}
	\hline
	\multicolumn{2}{c|}{ } & \multicolumn{3}{c}{ $m_{L_3}^2 + m_{H_u}^2(\text{AD} = 10^9 \gev)$ $[\tev]^2$} \\
	\hline
	$\tan\beta$ & $A_0$ [TeV] & FlexibleSUSY & First order & Zeroth order\\
	\hline
	\multirow{3}{*}{5} & 0 & $16.93$ & $15.69$ & $15.57$ \\
	& 5 & $16.67$& $15.29$& $15.64$\\
	& 10 & $10.62$ & $9.90$ & $9.46$ \\
	\hline
	\multirow{3}{*}{10} & 0 & $17.34$ & $16.12$ & $15.72$ \\
	& 5 & $17.07$ & $16.09$ & $15.07$ \\
	& 10 & $11.08$& $10.44$ & $9.91$ \\
	\hline	
	\multirow{3}{*}{20} & 0 & $17.27$ & $16.04$ & $ 15.87$ \\
	& 5 & $16.94$ & $16.02$ & $15.63$ \\
	& 10 & $10.72$ & $10.30$ & $8.58$ \\
	\hline
	\multirow{3}{*}{30} & 0 & $16.96$& $15.75$& $14.79$ \\
	& 5 & $16.50$ & $15.70$ & $14.80$ \\
	& 10 & $9.81$ & $10.83$ & $7.15$ \\
	\hline
	\multirow{3}{*}{40} & 0 & $16.40$ & $15.30$ & $13.03$\\
	& 5 & $15.74$ & $15.10$ & $13.41$ \\
	& 10 & $8.31$ & $8.73$ & $5.69$ \\
	\hline
\end{tabular}	
	\caption{$m_{L_3}^2 + m_{H_u}^2(\text{AD} = 10^9 \gev)$ from FlexibleSUSY, the zeroth and first order semi-analytic formulae for $m_0 = 3 \tev, M_{1/2} = 4\tev$ and $\mu >0$.}
	\label{mlhu2-value}
\end{table}

\noindent
To compare with the numerical package FlexibleSUSY, we reduce the parameter space to that of the constrained MSSM (CMSSM) which assumes a universal scalar mass, $m_0$, gaugino mass, $M_{1/2}$, and trilinear coupling, $A_0$ at the GUT scale, with $B$ and $|\mu|$ replaced by the electroweak symmetry breaking scale and $\tan\beta$. Table~\ref{mlhu2-value} shows the numerical values of $(m_{L_3}^2 + m^2_{H_u})(10^9\gev)$ from the following three methods
\begin{enumerate}[(1)]
\item the numerical package FlexibleSUSY 
\item the first order semi-analytic formulae, Eq.~\eqref{expansion1}
\item the zeroth order semi-analytic formulae, Eq.~\eqref{expansion0}
\end{enumerate}
assuming CMSSM boundary conditions with $m_0=3\tev, M_{1/2} = 4\tev, A_0 = 0, 5, 10\tev$ and $\tan\beta=5,10,20,30, 40$. For the two semi-analytic formulae, we took the low scale MSSM parameters ($D_t(0),A_t(0),\cdots.$) from FlexibleSUSY. 
Except for $\tan\beta=5$ and $A_0=5\tev$, the first order result is closer to the numerical result than the zeroth order result. Also, the difference between the first and zeroth order results increases as $\tan\beta$ increases as can be expected for the perturbative method.

\begin{figure}[htbp]
	\centering
	\includegraphics[width=0.9\linewidth]{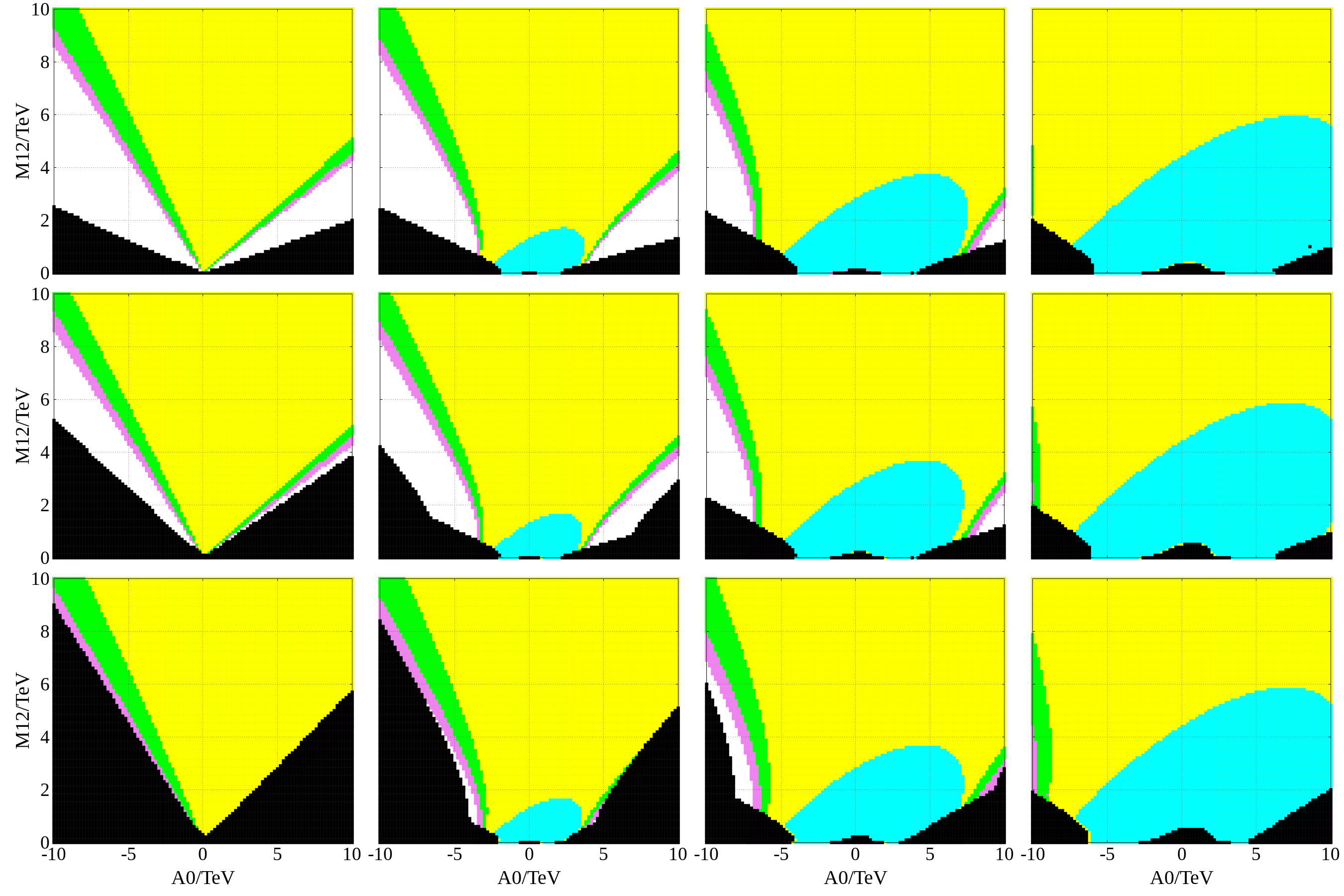}
	\caption{The CMSSM parameter space constrained by the baryogenesis condition and electroweak symmetry breaking. From top to bottom, $\tan\beta=10,20,30$. From left to right, $m_0=0,1,2,3\tev$. {\color[RGB]{0,255,255} $\blacksquare$} $\left.m_{L_3H_u}^2\right|_{m_s} > 0$ using FlexibleSUSY (violating the tree-level baryogenesis condition),  {\color[RGB]{0,255,0} $\blacksquare$} $\left.m_{L_3H_u}^2\right|_\text{AD} > 0$ using FlexibleSUSY (violating the renormalised baryogenesis condition), {\color[RGB]{255,255,0} $\blacksquare$} $\left.{m_{L_3H_u}^{2\,(0)}}_\text{S}\right|_\text{AD} > 0$ (violating the simplified zeroth order semi-analytic renormalised baryogenesis condition),  {\color[RGB]{238,130,238} $\blacksquare$} $\left.{m_{L_3H_u}^{2\,(0)}}\right|_\text{AD} > 0$ (violating Eq.~\eqref{gut formula}), $\blacksquare$ electroweak symmetry is unbroken or incorrectly broken.}
	\label{fig:comparison1}
\end{figure}

In Figure~\ref{fig:comparison1}, we plotted various versions of the baryogenesis condition on the CMSSM parameter space. Cyan corresponds to $m_{L_3H_u}^2 > 0$ at tree-level. Yellow and cyan regions are ruled out by the simplified zeroth order semi-analytic renormalised baryogenesis condition. Green, yellow, and cyan regions are ruled out by the numerically calculated renormalised baryogenesis condition. Purple, green, yellow and cyan regions are ruled out by the zeroth order baryogenesis condition given in Eq.~\eqref{gut formula}. From Eq.~\eqref{gut formula}, it is understandable that the slope of the boundary of the baryogenesis condition is straight for $m_0 = 0$ and curved for $m_0 \ne 0$. Black regions are ruled out by electroweak symmetry breaking. One cans see that electroweak symmetry breaking and the baryogenesis constraints are complementary. Meanwhile, the tree-level result is much weaker than the renormalised constraints, showing the importance of the renormalisation. Also, the renormalised constraints are similar, showing the robustness of the semi-analytic formulae.
\clearpage

\section{Constraining the MSCM}\label{s3}

\subsection{The cold dark matter abundance}
\noindent
In the MSCM of Eq.~\eqref{mscm}, the cold dark matter consists of axions \cite{Kawasaki:2014sqa,Klaer:2017ond} and axinos \cite{Covi:2009pq, Choi:2013lwa}. The decay temperature of the flaton after thermal inflation is (Eq.\,(94) in Ref.\,\cite{Kim:2008yu})
	\begin{align}
	T_d &\simeq 100\gev \left|\frac{m_A^2-|B|^2}{m_A^2}\right| \left(\frac{10^{11}\gev}{\phi_0}\right) \left(\frac{|\mu|}{\tev} \right)^2\left(\frac{10^2\gev}{m_\text{PQ}} \right)^{\frac{1}{2}}\left[\fn{f}{\frac{m_h^2}{m_\text{PQ}^2}} \right]^{\frac{1}{2}}\\
	&= \mathinner{10\gev} C_1  \left(\frac{ 10^{11}\gev}{\phi_0}\right) \left(\frac{|\mu|}{\tev} \right)^2 \label{Td expression} 
	\end{align}
where $C_1$ can be estimated as $\mathcal{O}(1)$ from Figure~6 in Ref.\,\cite{Kim:2008yu}.

\paragraph{Axion abundance}
For $T_d \gg 1\gev$, the axion abundance is \cite{Kawasaki:2014sqa,Klaer:2017ond}
	\begin{equation}\label{axion abundance}
	\Omega_a \simeq 0.1\left(\frac{f_a}{10^{11}\gev} \right)^{1.2}
	\end{equation}
$\phi_0$ can be expressed in terms of $f_a$,
	\begin{equation}\label{phi0}
	\phi_0 =\frac{N}{\sqrt{2}}f_a \simeq 4\times 10^{11}\gev \left(\frac{N}{6}\right)\left(\frac{f_a}{10^{11}\gev} \right)  
	\end{equation}
	where $N = \sum_i p_i$ and $p_i$ is the PQ charge of the $i$th quark.

\paragraph{Axino from flaton decay}
The effective interaction between the axino and the radial flaton is \cite{Kim:2008yu}
\begin{equation}
\frac{\alpha_{\tilde{a}}m_{\tilde{a}}}{\sqrt{2}\, \phi_0}\delta r\tilde{a}^2+\text{c.c.}
\end{equation}
where $m_{\tilde{a}}$ is the mass of the axino. Using the flaton to axinos decay rate
\begin{equation}
\Gamma_{\phi\rightarrow \tilde{a}\tilde{a}} = \frac{\alpha^2_{\tilde{a}}m^2_{\tilde{a}}m_\text{PQ}}{32\pi\phi_0^2}
\end{equation}
one can estimate the current abundance of axinos produced by the flaton decay as (Eq.\,(142) in Ref.\,\cite{Kim:2008yu})
\begin{equation}
\Omega_{\phi\rightarrow\tilde{a}} \simeq \mathinner{0.023} \frac{\Gamma_\phi^{1/2}}{\Gamma_\text{SM}^{1/2}}\left(\frac{100}{\fn{g_*}{T_d}} \right)^\frac{1}{2}\left(\frac{\alpha_{\tilde{a}}}{10^{-1}} \right)^2 \left(\frac{m_{\tilde{a}}}{\gev} \right)^3 \left(\frac{10\gev}{T_d} \right) \left(\frac{4\times 10^{11}\gev}{\phi_0} \right)^2
\end{equation}
where $\Gamma_\text{SM}$ is the rate of the flaton's decay to the Standard Model particles, and $\Gamma_\phi \simeq \Gamma_\text{SM}$ is the total decay rate of the flaton.

\paragraph{Axino from NLSP decay}
The NLSP decays to axinos with the decay rate
\begin{equation}
\Gamma_{N\rightarrow \tilde{a}} = \mathinner{A}\frac{m_N^3}{16\pi\phi_0^2}
\end{equation}
where $A\sim \mathcal{O}(1)$ and $m_N$ is the mass of the NLSP, producing the axino abundance (Eq.\,(166) in Ref.\,\cite{Kim:2008yu})
\begin{align}
\Omega_{N\rightarrow\tilde{a}} &\sim \mathinner{0.19} \mathinner{A}\frac{\Gamma^{1/2}_\text{SM}}{\Gamma_\phi^{1/2}}\left(\frac{10^3}{\fn{g_*}{T_d}^{3/2}\fn{g_*}{T_N}^{4/5}} \right) \left(\frac{m_N}{10^2\gev} \right) \left(\frac{m_{\tilde{a}}}{\gev} \right) \left(\frac{10^{11}\gev}{\phi_0} \right)^2\left(\frac{25T_d}{m_N} \right)^7\\
&= \mathinner{0.12} C_2 \left(\frac{m_N}{\tev} \right) \left(\frac{4\times 10^{11}\gev}{\phi_0} \right)^2\left(\frac{25T_d}{m_N} \right)^7 \label{abundance}
\end{align} 
where $C_2 \sim \mathcal{O}(1)$.

Assuming the axino abundance is less than the axion abundance
\begin{equation}\label{mN interm}
\mathinner{0.12}\mathinner{C_2} \left(\frac{m_N}{\tev} \right) \left(\frac{4\times 10^{11}\gev}{\phi_0} \right)^2\left(\frac{25T_d}{m_N} \right)^7 \lesssim \Omega_{\tilde{a}} \lesssim 0.1
\end{equation}
and substituting Eq.\,\eqref{Td expression} into Eq.\,\eqref{mN interm}, we get
	\begin{equation}
	m_N \gtrsim \mathinner{40\gev} C_1^{\frac{7}{6}}C_2^{\frac{1}{6}} \left(\frac{|\mu|}{\tev} \right)^{\frac{7}{3}} \left(\frac{4\times 10^{11}\gev}{\phi_0} \right)^{\frac{3}{2}}
	\end{equation}

\noindent
Referring to Eqs.\,\eqref{axion abundance} and \eqref{phi0}, we will use $f_a \lesssim 10^{11}\gev$ and hence $\phi_0 \lesssim 4\times 10^{11}\gev$ in the following discussion. Then, the mass of the NLSP reduces to
	\begin{equation}\label{dm}
	m_N \gtrsim 40\gev \left(\frac{|\mu|}{\tev}\right)^{\frac{7}{3}}
	\end{equation}

\subsection{Constraints}
\begin{figure}[htbp]
	\centering
	\includegraphics[width=.95\linewidth]{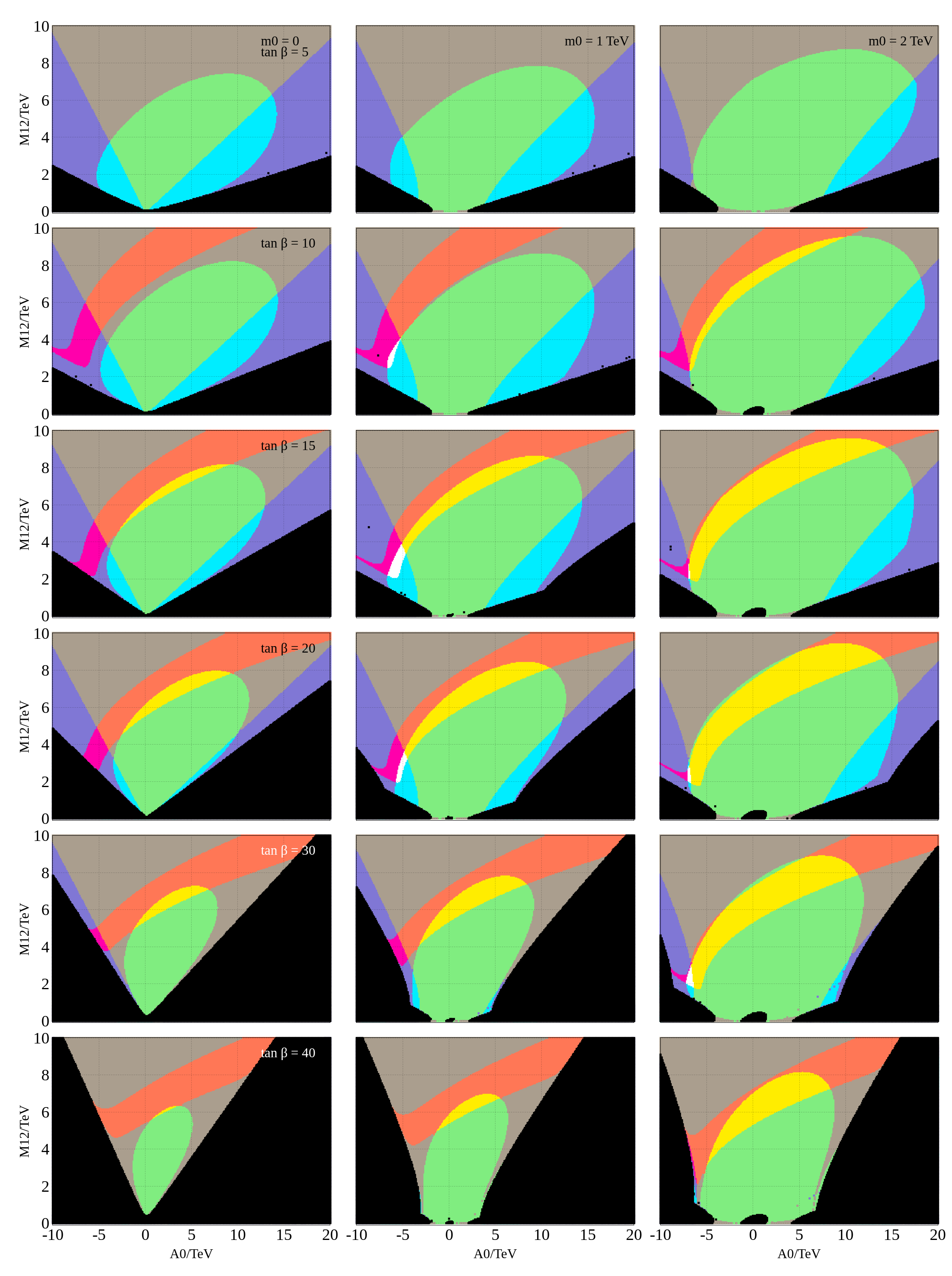}
	\caption{The CMSSM parameter space with four constraints. From top to bottom, $\tan\beta=5,10,15,20,30,40$. From left to right, $m_0=0,1,2\tev$. $\blacksquare$ electroweak symmetry is incorrectly broken,  {\color[RGB]{255,237,0} $\blacksquare$}  $\left.m_{L_3H_u}^2\right|_\text{AD}>0$ (i.e.\ baryogenesis constraint is violated), {\color[RGB]{0, 237, 255} $\blacksquare$} $m_h < 124.68\gev$ or $m_h > 125.68\gev$ (i.e.\ the Higgs mass constraint is violated), {\color[RGB]{255, 0, 171} $\blacksquare$} $m_N \lesssim 40\gev \left(\mu/\tev \right)^{\frac{7}{3}}$ (i.e.\ the axino dark matter constraint is violated). {\color[RGB]{128, 237, 128} $\blacksquare$} = {\color[RGB]{255,237,0}  $\blacksquare$}+{\color[RGB]{0, 237, 255}$\blacksquare$}, {\color[RGB]{255, 119, 86} $\blacksquare$} = {\color[RGB]{255,237,0}$\blacksquare$}+{\color[RGB]{255, 0, 171} $\blacksquare$}, {\color[RGB]{128, 119, 213} $\blacksquare$} = {\color[RGB]{0, 237, 255}$\blacksquare$}+{\color[RGB]{255, 0, 171} $\blacksquare$}, {\color[RGB]{170, 158, 142} $\blacksquare$} = {\color[RGB]{255,237,0}$\blacksquare$}+{\color[RGB]{0, 237, 255}$\blacksquare$}+{\color[RGB]{255, 0, 171}$\blacksquare$}. White is the only allowed region satisfying all constraints.}
	\label{fig:fsmlhu2mhnlsp-1}
\end{figure}

\begin{figure}[htbp]
	\centering
	\includegraphics[width=.95\linewidth]{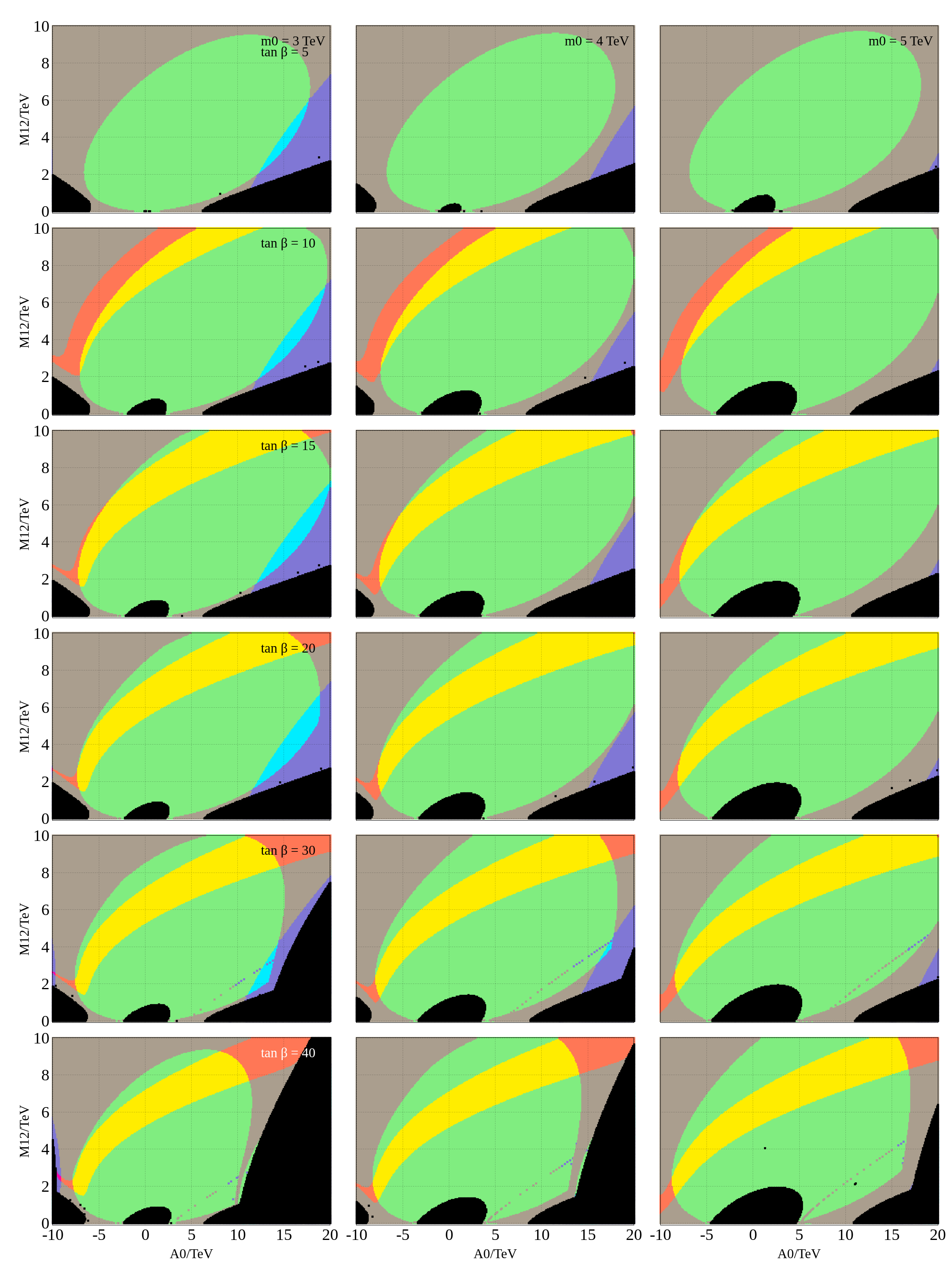}
	\caption{CMSSM parameter space with four constraints. From top to bottom, $\tan\beta=5,10,15,20,30,40$. From left to right, $m_0=3,4,5\tev$. $\blacksquare$ electroweak symmetry is incorrectly broken,  {\color[RGB]{255,237,0} $\blacksquare$}  $\left.m_{L_3H_u}^2\right|_\text{AD}>0$ (i.e.\ baryogenesis constraint is violated), {\color[RGB]{0, 237, 255} $\blacksquare$} $m_h < 124.68\gev$ or $m_h > 125.68\gev$ (i.e.\ the Higgs mass constraint is violated), {\color[RGB]{255, 0, 171} $\blacksquare$} $m_N \lesssim 40\gev \left(\mu/\tev \right)^{\frac{7}{3}}$ (i.e.\ the axino dark matter constraint is violated). {\color[RGB]{128, 237, 128} $\blacksquare$} = {\color[RGB]{255,237,0}  $\blacksquare$}+{\color[RGB]{0, 237, 255}$\blacksquare$}, {\color[RGB]{255, 119, 86} $\blacksquare$} = {\color[RGB]{255,237,0}$\blacksquare$}+{\color[RGB]{255, 0, 171} $\blacksquare$}, {\color[RGB]{128, 119, 213} $\blacksquare$} = {\color[RGB]{0, 237, 255}$\blacksquare$}+{\color[RGB]{255, 0, 171} $\blacksquare$}, {\color[RGB]{170, 158, 142} $\blacksquare$} = {\color[RGB]{255,237,0}$\blacksquare$}+{\color[RGB]{0, 237, 255}$\blacksquare$}+{\color[RGB]{255, 0, 171}$\blacksquare$}. White is the only allowed region satisfying all constraints.}
	\label{fig:fsmlhu2mhnlsp-2}
\end{figure}

In Figure~\ref{fig:fsmlhu2mhnlsp-1} and \ref{fig:fsmlhu2mhnlsp-2}, we imposed the following constraints on the CMSSM parameters using FlexibleSUSY
\begin{enumerate}[(1)]
	\item Electroweak symmetry breaking
	\item Baryogenesis condition
	\begin{equation}
	\left.m^2_{L_3H_u}\right|_\text{AD} < 0
	\end{equation}	
	\item Higgs mass 
	\begin{equation} \label{Mh}
	124.68\gev < m_h < 125.68\gev
	\end{equation}
	\item Axino dark matter abundance
	\begin{equation}
	m_N \gtrsim 40\gev \left(\frac{|\mu|}{\tev}\right)^{\frac{7}{3}} \tag{\ref{dm}}
	\end{equation}
\end{enumerate}

The dark matter constraint allows parameters inside the  white, yellow, cyan and green (i.e.\ colors excluding magenta) oval in the middle. For $\tan\beta \ge 10$, the Higgs mass constraint is satisfied inside the white, yellow, magenta and red (i.e.\ colors excluding cyan) tick ($\checkmark$) extending from the bottom left to the top right. The baryogensis condition is satisfied in the white, cyan, magenta and purple (i.e. colors excluding yellow) bands next to the regions prohibited by the electroweak symmetry breaking constraint. All four constraints are satisfied in the white overlapping regions.



\subsection{$H_u H_d$ constraint} \label{rhn}
Since $m_{L_3}^2 > 0$, the baryogenesis condition
\begin{equation}
\left.m_{L_3H_u}^2\right|_\text{AD} < 0 \tag{\ref{bg condition}}
\end{equation}
is achieved by having $m_{H_u}^2<0$. However, this may cause $H_uH_d$ to become large instead of $LH_u$. To avoid this, we require 
\begin{equation}\label{mhuhd}
\left. m_{H_uH_d}^2 \right|_\text{AD} > 0
\end{equation}
 
The renormalisation group equations relevant to these conditions are
\begin{eqnarray}
\frac{d}{dx}m_{L_3}^2 &=& \lambda_\tau^2A_\tau^2 \purple{+\lambda_\nu^2A_\nu^2}-3g_2^2M_2^2 - \frac{3}{5}g_1^2M_1^2 - \frac{3}{10}g_1^2\mathcal{S}+\lambda_\tau^2D_\tau \purple{+\lambda_\nu^2D_\nu}\label{RGE:ml32}\\
\frac{d}{dx}m_{H_u}^2 &=& 3\lambda_t^2A_t^2 \purple{+\lambda_\nu^2A_\nu^2} -3g_2^2M_2^2 - \frac{3}{5}g_1^2M_1^2 + \frac{3}{10}g_1^2\mathcal{S}+3\lambda_t^2D_t^2 \purple{+\lambda_\nu^2D_\nu} \label{RGE:mhu2}\\
\frac{d}{dx}(m_{H_d}^2-m_{L_3}^2) &=& 3\lambda_b^2(A_b^2+D_b) \purple{- \lambda_\nu^2(A_\nu^2+D_\nu)} \label{RGE:mhd-l}
\end{eqnarray}
where \purple{purple} terms are from right-handed neutrinos.
Eq.~\eqref{RGE:mhd-l} implies that $m_{H_uH_d}^2 < m_{L_3H_u}^2$ in the CMSSM without right-handed neutrinos. Hence, $m_{L_3H_u}^2 <0 $ and $m_{H_uH_d}^2 > 0$ cannot both be satisfied. However, with right-handed neutrinos \purple{(purple)} in Eq.\,\eqref{RGE:mhd-l}, both $m_{L_3H_u}^2 <0 $ and $m_{H_uH_d}^2 > 0$ can simultaneously be satisfied. Therefore, to avoid the instability along the $H_uH_d$ direction, one should relax the CMSSM boundary conditions or include right-handed neutrinos to the theory.

We adopt the $\nu$CMSSM~\cite{Kadota:2009fg} to consider the effect of right-handed neutrinos from $M_\text{GUT}$ to $M_{\bar\nu}$ and integrated them out below $M_{\bar\nu}$. Because $M_{\bar\nu}$ is close to $M_\text{GUT}$, we simply integrated their effects linearly in Eqs.~\eqref{RGE:ml32} and \eqref{RGE:mhu2}. Namely, we modified the CMSSM boundary conditions for $m_{L_3}^2$ and $m_{H_u}^2$ as
		\begin{align}
		m_{L_3}^2, m_{H_u}^2 &= m_0^2 \rightarrow m_0^2+ \frac{\lambda_\nu^2}{8\pi^2}\left(A_0^2+3m_0^2 \right)\log\frac{M_{\bar\nu}}{M_\text{GUT}} 
		\end{align}
and generated the low scale parameters using FlexibleSUSY. We also assumed $M_{\bar\nu} = 10^{14}\gev$ and $\lambda_\nu = \lambda_t$ at $M_\text{GUT} = 1.24\times 10^{16}\gev$. Meanwhile, we neglected the effects of right-handed neutrinos on the Yukawa and trilinear couplings.

\begin{figure}[htbp]
	\centering
	\includegraphics[width=\linewidth]{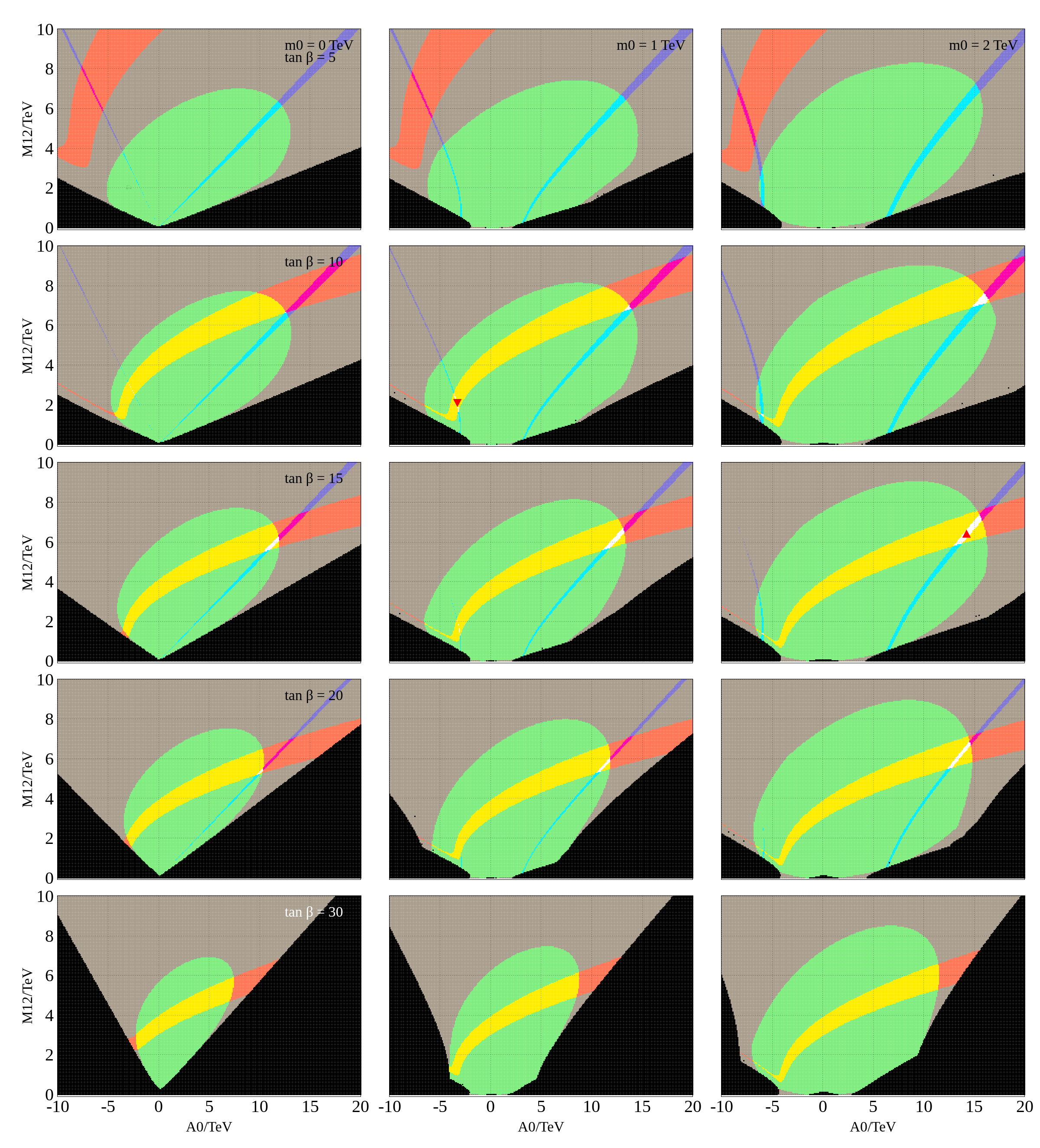}
	\caption{The CMSSM parameter space with four constraints in the presence of the right-handed neutrino. From top to bottom, $\tan\beta=5,10,15,20,30$. From left to right, $m_0=0,1,2\tev$. $\blacksquare$ electroweak symmetry is incorrectly broken, {\color[RGB]{255,237,0} $\blacksquare$}  $\left.m_{L_3H_u}^2\right|_\text{AD} > 0$ or $\left.m_{H_uH_d}^2\right|_\text{AD} < 0$ (i.e.\ baryogenesis constraint is violated), {\color[RGB]{0, 237, 255} $\blacksquare$} $m_h < 124.68\gev$ or $m_h > 125.68\gev$ (i.e.\ Higgs mass constraint is violated), {\color[RGB]{255, 0, 171} $\blacksquare$} $m_N \lesssim 40\gev \left(\mu/\tev \right)^{\frac{7}{3}}$ (i.e.\ the axino dark matter constraint is violated). {\color[RGB]{128, 237, 128} $\blacksquare$} = {\color[RGB]{255,237,0}  $\blacksquare$}+{\color[RGB]{0, 237, 255}$\blacksquare$}, {\color[RGB]{255, 119, 86} $\blacksquare$} = {\color[RGB]{255,237,0}$\blacksquare$}+{\color[RGB]{255, 0, 171} $\blacksquare$}, {\color[RGB]{128, 119, 213} $\blacksquare$} = {\color[RGB]{0, 237, 255}$\blacksquare$}+{\color[RGB]{255, 0, 171} $\blacksquare$}, {\color[RGB]{170, 158, 142} $\blacksquare$} = {\color[RGB]{255,237,0}$\blacksquare$}+{\color[RGB]{0, 237, 255}$\blacksquare$}+{\color[RGB]{255, 0, 171}$\blacksquare$}. White is the only allowed region satisfying all constraints. A spectrum for each central value is given at Figure~\ref{fig:spectrum} and Tables~\ref{spectrum-l} and \ref{spectrum-r}.}
	\label{fig:rh-1}
\end{figure}

\begin{figure}[htbp]
	\centering
	\includegraphics[width=\linewidth]{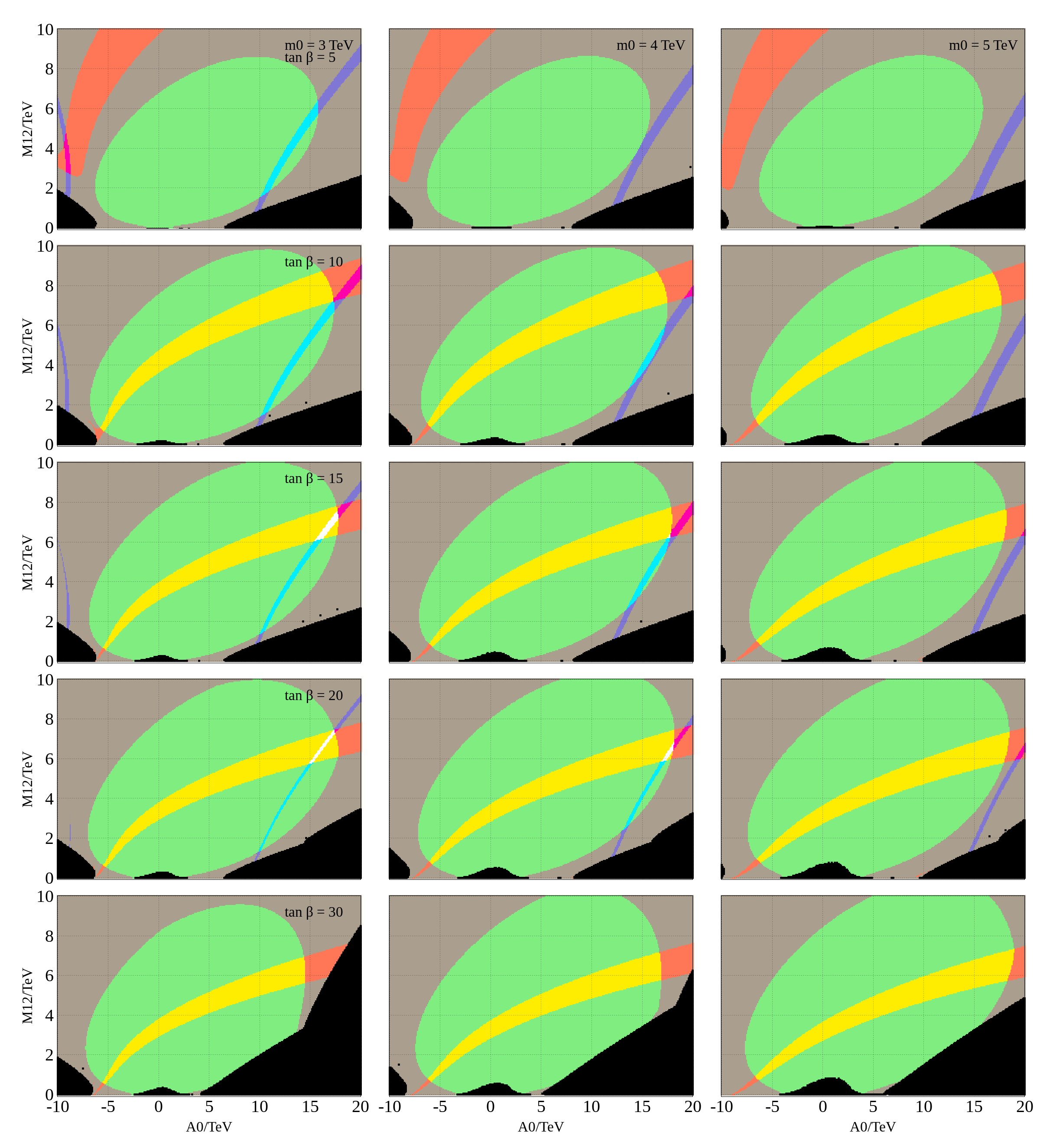}
	\caption{The CMSSM parameter space with four constraints in the presence of the right-handed neutrino. From top to bottom, $\tan\beta=5,10,15,20,30$. From left to right, $m_0=3,4,5\tev$. $\blacksquare$ electroweak symmetry is incorrectly broken, {\color[RGB]{255,237,0} $\blacksquare$}  $\left.m_{L_3H_u}^2\right|_\text{AD} > 0$ or $\left.m_{H_uHd}^2\right|_\text{AD} < 0$ (i.e.\ baryogenesis constraint is violated), {\color[RGB]{0, 237, 255} $\blacksquare$} $m_h < 124.68\gev$ or $m_h > 125.68\gev$ (i.e.\ Higgs mass constraint is violated), {\color[RGB]{255, 0, 171} $\blacksquare$} $m_N \lesssim 40\gev \left(\mu/\tev \right)^{\frac{7}{3}}$ (i.e.\ the axino dark matter constraint is violated). {\color[RGB]{128, 237, 128} $\blacksquare$} = {\color[RGB]{255,237,0}  $\blacksquare$}+{\color[RGB]{0, 237, 255}$\blacksquare$}, {\color[RGB]{255, 119, 86} $\blacksquare$} = {\color[RGB]{255,237,0}$\blacksquare$}+{\color[RGB]{255, 0, 171} $\blacksquare$}, {\color[RGB]{128, 119, 213} $\blacksquare$} = {\color[RGB]{0, 237, 255}$\blacksquare$}+{\color[RGB]{255, 0, 171} $\blacksquare$}, {\color[RGB]{170, 158, 142} $\blacksquare$} = {\color[RGB]{255,237,0}$\blacksquare$}+{\color[RGB]{0, 237, 255}$\blacksquare$}+{\color[RGB]{255, 0, 171}$\blacksquare$}. White is the only allowed region satisfying all constraints.}
	\label{fig:rh-2}
\end{figure}	

Extending the baryogenesis condition to the combination of Eqs.~\eqref{bg condition} and \eqref{mhuhd}, we used the same four constraints---the baryogenesis condition, electroweak symmetry breaking, Higgs mass and the axino dark matter abundance---to constrain the CMSSM parameter space. Figures~\ref{fig:rh-1} and \ref{fig:rh-2} show the resulting CMSSM parameter space.
\begin{figure}
	\centering
	\begin{subfigure}[htbp]{0.44\textwidth}
		\centering
		\includegraphics[width=\textwidth]{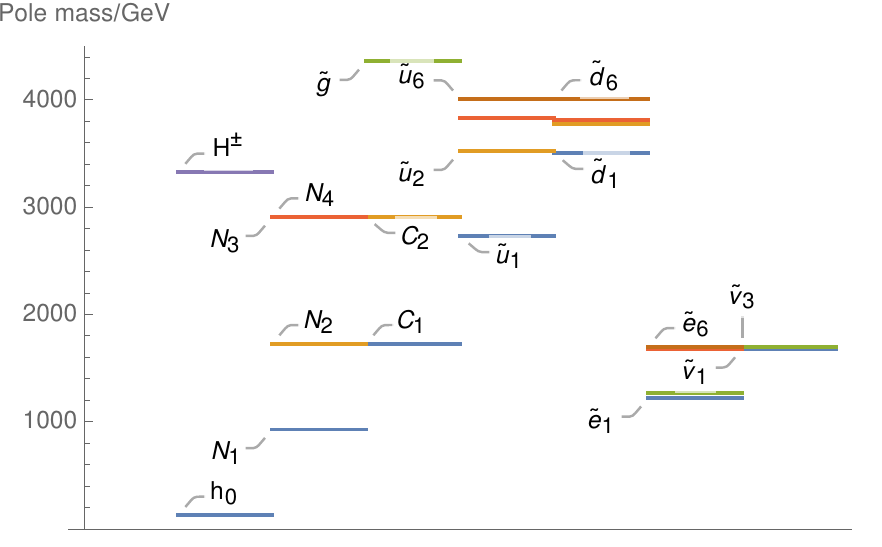}
		\caption{Left-hand side region}
		\label{fig:spectrum-l}
	\end{subfigure}
	\hfill
	\begin{subfigure}[htbp]{0.44\textwidth}
		\centering
		\includegraphics[width=\textwidth]{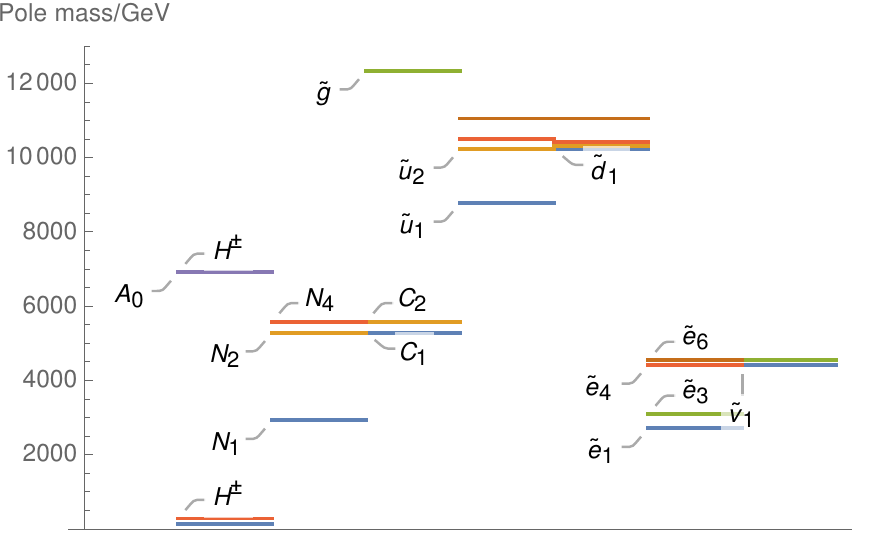}
		\caption{Right-hand side region}
		\label{fig:spectrum-r}
	\end{subfigure}
	\caption{Sparticle spectra for the central points on the left- and right-hand side regions satisfying all four constraints in Figure~\ref{fig:rh-1}. Higgs scalar mass eigenstates consist of $h^0, H^0$: CP even neutral scalars, $A^0$: CP odd neutral scalar, $H^{\pm}$: CP odd charge +1 scalars. $\tilde{g}$: gluino, $N_i$: neutralinos, $C_j$: charginos, $\tilde{u}_k$: up-type squarks, $\tilde{d}_k$: down-type squarks, $\tilde{e}_k$: sleptons, $\tilde{\nu}_l$: sneutrinos where $i=1,2,3,4$, $j=1,2$, $k=1,\cdots,6$, $l=1,2,3$, and each index is in ascending order of pole mass.}
	\label{fig:spectrum}
\end{figure}

\begin{table}[htbp]
	\centering
	\setlength\extrarowheight{3pt}
	\begin{tabular}{c|c|c|c||c|c|c|c|c|c}
		\hline
		$h^0$ & $H^0$ & $A^0$ & $H^{\pm}$ & $\tilde{u}_1$ & $\tilde{u}_2$ & $\tilde{u}_3$ & $\tilde{u}_4$ & $\tilde{u}_5$ & $\tilde{u}_6$ \\
		\hline
		0.1253 & 3.323 & 3.323 & 3.324 & 2.728 & 3.520 & 3.832 & 3.832 & 4.007 & 4.007 \\
		\hline
		\hline
		$N_1$ & $N_2$ & $N_3$ & $N_4$ & $\tilde{d}_1$ & $\tilde{d}_2$ & $\tilde{d}_3$ & $\tilde{d}_4$ & $\tilde{d}_5$ & $\tilde{d}_6$\\
		\hline
		0.9255 & 1.726 & 2.904 & 2.906 & 3.503 & 3.777 & 3.811 & 3.811 & 4.008 & 4.008\\
		\hline
		\hline
		$C_1$ & $C_2$ & $\tilde{g}$ & & $\tilde{e}_1$ & $\tilde{e}_2$ & $\tilde{e}_3$ & $\tilde{e}_4$ & $\tilde{e}_5$ & $\tilde{e}_6$\\
		\hline
		1.726 & 2.907 & 4.538 & & 1.219 & 1.269 & 1.269 & 1.680 & 1.697 & 1.697 \\
		\hline
		\hline
		& & & &  $\tilde{\nu}_1$ & $\tilde{\nu}_2$ & $\tilde{\nu}_3$ & & & \\
		\hline
		& & & & 1.677 & 1.695 & 1.695 & & & \\
		\hline
	\end{tabular}
	\caption{$\tev$ scale pole masses of sparticles for the central point on the left-hand side region satisfying all four constraint which corresponds to $\tan\beta = 10, m_0 = 1 \tev, M_{1/2}=2.1\tev,A_0 = -3.3\tev$. }
	\label{spectrum-l}
\end{table}

\begin{table}[htbp]
	\centering
	\setlength\extrarowheight{3pt}
	\begin{tabular}{c|c|c|c||c|c|c|c|c|c}
		\hline
		$h^0$ & $H^0$ & $A^0$ & $H^{\pm}$ & $\tilde{u}_1$ & $\tilde{u}_2$ & $\tilde{u}_3$ & $\tilde{u}_4$ & $\tilde{u}_5$ & $\tilde{u}_6$ \\
		\hline
		0.1252 & 7.034 & 7.034 & 7.035 & 8.966 & 10.45 & 10.71 & 10.71 & 11.28 & 11.28 \\
		\hline
		\hline
		$N_1$ & $N_2$ & $N_3$ & $N_4$ & $\tilde{d}_1$ & $\tilde{d}_2$ & $\tilde{d}_3$ & $\tilde{d}_4$ & $\tilde{d}_5$ & $\tilde{d}_6$\\
		\hline
		2.995 & 5.406 & 5.666 & 5.679 & 10.45 & 10.54 & 10.63 & 10.63 & 11.28 & 11.28\\
		\hline
		\hline
		$C_1$ & $C_2$ & $\tilde{g}$ & & $\tilde{e}_1$ & $\tilde{e}_2$ & $\tilde{e}_3$ & $\tilde{e}_4$ & $\tilde{e}_5$ & $\tilde{e}_6$\\
		\hline
		5.406 & 5.679 & 12.61 & & 2.745 & 3.125 & 3.126 & 4.495 & 4.619 & 4.619 \\
		\hline
		\hline
		& & & &  $\tilde{\nu}_1$ & $\tilde{\nu}_2$ & $\tilde{\nu}_3$ & & & \\
		\hline
		& & & & 4.494 & 4.617 & 4.618 & & & \\
		\hline
	\end{tabular}
	\caption{$\tev$ scale pole masses of sparticles for the central point on the right-hand side region satisfying all four constraint which corresponds to $\tan\beta = 15, m_0=2\tev,M_{1/2}=6.55\tev,A_0=14.45\tev$. }
	\label{spectrum-r}
\end{table}

From Figures~\ref{fig:rh-1} and \ref{fig:rh-2}, $\left.m_{H_uH_d}^2\right|_\text{AD}>0$ places a bound complementary to $\left.m_{L_3H_u}^2\right|_\text{AD} < 0$ on the parameter space. This leaves only a thin strip along the boundary of the previous baryogenesis condition, making the combined baryogenesis condition very restrictive. In contrast, Eq.\,\eqref{RGEmlhu} shows that the presence of the right-handed neutrino drives $m_{L_3H_u}^2$ more negative at low scales including AD scale, relaxing the previous baryogenesis condition ($\left.m_{L_3H_u}^2 \right|_\text{AD} < 0$). Consequently, there are two regions satisfying all four constraints on the left- and right-hand side. Central points of each region are maked by inverted triangle (left) and triangle (right) in Figure~\ref{fig:rh-1}. Figure~\ref{fig:spectrum} shows the spectrum of sparticles in each case, and corresponding pole masses are given at Tables~\ref{spectrum-l} and \ref{spectrum-r}.


\section{Conclusion}

Affleck-Dine leptogenesis after thermal inflation along the $LH_u$ direction \cite{Stewart:1996ai,Jeong:2004hy,Kawasaki:2006py,Felder:2007iz,Kim:2008yu} requires $m_{LH_u}^2 < 0$ up to the AD scale ($|L|\simeq|H_u|\sim 10^9\gev$). 
In Section \ref{s2}, we renormalised this baryogenesis condition from the AD scale to the soft supersymmetry breaking scale by solving the renormalisation group equations perturbatively in the Yukawa couplings. The resulting zeroth and first order semi-analytic constraints on the soft supersymmetry breaking parameters are Eqs.\,\eqref{expansion0} and \eqref{expansion1} with the numerical coefficients given in Tables \ref{tanb5}--\ref{tanb30}. Since the $M_1$ contributions are small, we set $M_1=0$ to get the simplified zeroth order formula given in Eq.\,\eqref{simplified formula}. The robustness of our formula can be seen in several ways. Firstly, the numerical coefficients in the zeroth and first order formulae are close. Also, in Table \ref{mlhu2-value}, the numerical values of $m_{L_3H_u}^2$ at the AD scale from the semi-analytic formulae and the numerical package FlexibleSUSY are fairly close. Lastly, in Figure~\ref{fig:comparison1}, the simplified zeroth order formula constrains the CMSSM parameter space similarly to the fully numerical result but much stronger than the tree-level formula.

In Section \ref{s3}, we used the numerical package FlexibleSUSY to renormalise the baryogenesis condition assuming CMSSM boundary conditions. We considered the MSCM of Eq.\,\eqref{mscm} and combined the renormalised baryogenesis condition with other constraints, specifically, electroweak symmetry breaking, Higgs mass, Eq.\,\eqref{Mh}, and the cold dark matter abundance, Eq.\,\eqref{dm}. In Figures~\ref{fig:fsmlhu2mhnlsp-1} and \ref{fig:fsmlhu2mhnlsp-2}, there is a region satisfying all the four constraints, which corresponds to 
\begin{equation}
10\lesssim \tan\beta \lesssim 30,\quad 1\tev \lesssim m_0 \lesssim 2\tev, \quad 2\tev \lesssim M_{1/2} \lesssim 4\tev, \quad -7.5\tev \lesssim  A_0 \lesssim -5 \tev
\end{equation}


In Section \ref{rhn}, we considered the stability of the $H_uH_d$ direction by adding the additional constraint $\left.m_{H_uH_d}^2\right|_\text{AD}>0$, which is incompatible with $\left.m_{L_3H_u}^2\right|_\text{AD}<0$ if we assume CMSSM boundary conditions and neglect the effects of right-handed neutrinos. Including the effects of right-handed neutrinos, $\left.m_{H_uH_d}^2\right|_\text{AD}>0$ and $\left.m_{L_3H_u}^2\right|_\text{AD}<0$ constrain the CMSSM parameter space to narrow bands. Combining this baryogenesis condition with the other constraints in Figures~\ref{fig:rh-1} and \ref{fig:rh-2}, there are two regions satisfying all four constraints. One on the left-hand side corresponding to 
\begin{equation}
10\lesssim \tan\beta \lesssim 20,\quad 0 \lesssim m_0 \lesssim 2\tev, \quad 1.5\tev \lesssim M_{1/2} \lesssim 3\tev, \quad -6\tev \lesssim A_0 \lesssim -2.5\tev
\end{equation}
and the other on the right-hand side corresponding to 
\begin{equation}
10\lesssim \tan\beta \lesssim 20, \quad 0\lesssim m_0 \lesssim 4\tev, \quad 5\tev \lesssim M_{1/2} \lesssim 7.5\tev, \quad 10\tev \lesssim A_0 \lesssim 18\tev
\end{equation}

Lastly, we suggest to add an extra constraint---tunneling to non-MSSM vacua---for the future work. The baryogenesis condition, $m_{L_3H_u}^2<0$ below AD scale, implies non-MSSM deeper vacua ($L,H_u,Q,\bar{d}$) or ($L,H_u,\bar{e}$) \cite{Komatsu:1988mt,Casas:1995pd} to which the MSSM vacuum can tunnel to. Requiring the life time of the MSSM vacuum to be larger than the age of the Universe would constrain the CMSSM parameters in a manner complementary to the baryogenesis condition and similar to but stronger than the electroweak symmetry breaking condition.

\section*{Acknowledgements}
\addcontentsline{toc}{section}{Acknowledgement}
\noindent
EDS thanks Gabriela Barenboim, Wan-il Park and Javier Rasero for collaboration on an earlier version of this project and Kenji Kadota for helpful discussions. SK thanks Jae-hyeon Park for helping with FlexibleSUSY. We thank Wan-il Park for helpful comments on a draft of this paper.

\begin{appendices}
\section{Field-dependent renormalisation using RG improvement} \label{rgimprovement}

The renormalization group improved potential satisfies \cite{Callan}
	\begin{equation}
	\mu \frac{d}{d\mu}V (\mu,\lambda_i,\phi) = \left (\mu\frac{\partial}{\partial \mu} + \beta_{\lambda_i}\frac{\partial}{\partial \lambda_i} - \gamma\phi \frac{\partial}{\partial \phi} \right) V (\mu,\lambda_i,\phi) = 0 \label{rge}
	\end{equation}
where $\beta_{\lambda_i}$ are beta functions of the couplings $\lambda_i$ and $\gamma$ is the wavefunction renormalization. One can solve the Eq.\,\eqref{rge} using the method of characteristic \cite{Ford:1992mv}. This method regards each variable $\mu, \lambda$ and $\phi$ as points on a curve parametrized by $t$,
	\begin{equation}
	\begin{split}
	V (\mu,\lambda,\phi) &= V (\mu (t),\lambda (t),\phi (t))
	\end{split}
	\end{equation} 
where the variables satisfy
	\begin{equation}
	\frac{d\mu}{dt}= \mu, \quad \frac{d\lambda_i}{dt} = \beta_{\lambda_i}, \quad \frac{d\phi}{dt} = -\gamma\phi
	\end{equation}
with the solution 
	\begin{equation}\label{vareq}
	\mu (t) = \mu_0 e^t, \quad \phi (t) = \phi(0) \exp\left (-\int^t_0 \gamma (\lambda_i (t'))dt' \right)
	\end{equation}
	
Then, the renormalization scale, field values and couplings are functions of $t$ such that any of their changes with respect to the $t$ cancel each other so that the potential is left invariant. Hence, if the inverse map of Eq.\,\eqref{vareq} exists, one can relate the field value to the renormalization scale. 
Any choice of $t$ allows one to connect the renormalization scale to the field value. However, there is some choice of $t$ that simplifies the RG improved potential in 1-loop order. For example, consider a 1-loop effective potential
\begin{equation}
V_{1-\text{loop}} = \frac{1}{64\pi^2}\text{STr}\left (M^4 (\lambda_i (t),\phi (t))\ln\frac{M^2 (\lambda_i (t),\phi (t))}{\mu_0^2 e^{2t}} -\frac{3}{2}\right)
\end{equation}
with the dominant eigenvalues of $M^2$ are close to each other. Let $\bar{M}^2$ denotes one of the dominant eigenvalues of $M^2$ and choose $t$
	\begin{equation}\label{t-cancel-large-log}
t = \frac{1}{2}\ln\frac{\bar{M}^2 (\lambda_i (t),\phi (t))}{\mu_0^2}
\end{equation}
then it follows that
	\begin{equation}
\begin{split}
V (\mu,\lambda,\phi) &= V_\text{tree} (\mu (t),\lambda (t),\phi (t)) + \text{subleading terms}
\end{split}
\end{equation} 
Thus, with the choice of $t$ in Eq.\,\eqref{t-cancel-large-log}, the RG improved 1-loop effective potential reduces to the tree-level potential with renormalized variables. Moreover, this choice of $t$ manifests the field dependent renormalization, i.e. $\phi \rightarrow t(\phi) \rightarrow \lambda_i(\phi)$, through the implicit $t$-dependence. Since one can find the renormalization scale $\mu(t)$ corresponding to the $t$, the renormalization of the couplings with respect to the renormalization scale leads to the renormalization of the couplings with respect to the field value.

\section{Perturbative solution of the renormalisation group equations}\label{perturbation}

\subsection{Zeroth order solution}\label{a2-1}
In this Appendix, we solve Eq.\,\eqref{RGEmlhu} analytically neglecting colored terms to obtain Eq.\,\eqref{mlhu0}. Substituting Eq.\,\eqref{RGEmlt} into Eq.\,\eqref{RGEmlhu} gives
\begin{equation}
\frac{d}{dx} \left(m_{L_3}^2 + m_{H_u}^{2}\right)^{(0)} = \frac{1}{2}\frac{d}{dx}D_t^{(0)} + \frac{16}{3}g_3^2M_3^2 -3g_2^2M_2^2 -\frac{1}{3}g_1^2M_1^2
\end{equation}
which can be solved as
\begin{equation}
\left[m_{L_3}^{2}(x)+m_{H_u}^{2}(x) \right]^{(0)} = m_{L_3}^2(0) + m_{H_u}^2(0) + \frac{1}{2}\left[D_t^{(0)}(x)-D_t(0)\right] - \int^x_0 \left(\frac{16}{3}g_3^2M_3^2 - 3g_2^2M_2^2 - \frac{1}{3}g_1^2M_1^2 \right) \hspace{0.1cm} dt \label{RGE:mlhu0}
\end{equation}
To evaluate (\ref{RGE:mlhu0}), it is enough to solve for $D_t$. Solving Eqs.\,\eqref{RGElt},\eqref{RGEAt} and \eqref{RGEmlt} using Eq.\,\eqref{formula2},
\begin{align}
\frac{1}{\lambda_t^{2(0)}} & = e^{\int^x_0 \left(\frac{16}{3}g_3^2 + 3g_2^2+\frac{13}{15}g_1^2\right) \dt} \left[\frac{1}{\lambda_t^2(0)} -6\int^x_0 e^{-\int^t_0 \left(\frac{16}{3}g_3^2 + 3g_2^2+\frac{13}{15}g_1^2 \right) \dt'} \dt \right] \label{Sollt}\\
A_t(x)^{(0)} &= e^{\int^x_0 6\lambda_t^{2(0)} \dt}\left[A_t(0) + \int^x_0 e^{-\int^t_06\lambda_t^{2(0)} \dt'}\left(\frac{16}{3}g_3^2M_3^2 + 3g_2^2M_2^2 +\frac{13}{15}g_1^2M_1^2 \right)\dt \right] \label{SolAt}\\
D_t(x)^{(0)} &= e^{\int^x_0 6\lambda_t^{2(0)} \dt}\left[D_t(0) + \int^x_0 e^{-\int^t_06\lambda_t^{2(0)} \dt'}\left(6\lambda_t^{2(0)}A_t^{2(0)} -\frac{32}{3}g_3^2M_3^2 - 6 g_2^2M_2^2 -\frac{26}{15}g_1^2M_1^2 \dt\right)\right]\label{SolDt}
\end{align}
Using Eq.\,\eqref{SolAt} and integrating by parts,
\begin{align}
\nonumber \int^x_0 &e^{-\int^t_0 6\lambda_t^{2(0)} \dt} 6\lambda_t^{2(0)} A_t^2 \dt\\
&= \int^x_0 6\lambda_t^{2(0)} e^{\int^t_0 6\lambda_t^{2(0)}\dt'}
\left[A_t(0) + \int^t_0 e^{-\int 6\lambda_t^{2(0)} \dt'}
\left(\frac{16}{3}g_3^2M_3 + 3g_2^2M_2 + \frac{13}{15}g_1^2M_1 \right)\dt \right]^2\\
\nonumber &=-A_t^2(0) + e^{\int^x_0 {6\lambda_t^{2(0)}}\dt}
\left[A_t(0) + \int^t_0 e^{-\int^t_0 6\lambda_t^{2(0)}\dt'}
\left(\frac{16}{3}g_3^2M_3 + 3g_2^2M_2 + \frac{13}{15}g_1^2M_1 \dt\right)\right]^2\\
& \quad -2\int^x_0 \left(\frac{16}{3}g_3^2M_3 + 3g_2^2M_2 + \frac{13}{15}g_1^2M_1\right)
\left[A_t(0) + \int^t_0 e^{-\int^{t'}_0 6\lambda_t^{2(0)} \dt''}
\left(\frac{16}{3}g_3^2M_3 + 3g_2^2M_2 + \frac{13}{15}g_1^2M_1\right)\dt'\right]dt \label{intermediate1}
\end{align}
Using Eqs\,\eqref{Solg1} to \eqref{SolM3},
\begin{equation}
\int \left(\frac{16}{3}g_3^2M_3 + 3g_2^2M_2 + \frac{13}{15}g_1^2M_1\right) \hspace{1mm} dx = -\frac{16}{9}M_3(x) + 3M_2(x) +\frac{13}{99}M_1(x) \label{technic1}
\end{equation}
and hence Eq.\,\eqref{intermediate1} can be integrated by parts,
\begin{align}
\nonumber \int^x_0  &e^{-\int^t_0 6\lambda_t^{2(0)} \dt} 6\lambda_t^{2(0)}A_t^{2(0)} \dt\\
\nonumber &= -A_t^2(0) -2A_t(0)\int^x_0 \left(\frac{16}{3}g_3^2M_3 + 3g_2^2 + \frac{13}{15}g_1^2M_1\right) \dt \\
\nonumber &\quad {} + e^{\int 6\lambda_t^{2(0)} \dt}\left[A_t(0) + \int^x_0 e^{-\int^t_0 6\lambda_t^{2(0)}\dt'}
\left(\frac{16}{3}g_3^2M_3 + 3g_2^2M_2 + \frac{13}{15}g_1^2M_1\right) \dt \right]^2\\
\nonumber &\quad {}+2\left(\frac{16}{9}M_3 - 3M_2 - \frac{13}{99}M_1\right)
\int^x_0 e^{-\int^t_0 6\lambda_t^{2(0)}\dt'}\left(\frac{16}{3}g_3^2M_3 + 3g_2^2M_2 + \frac{13}{15}g_1^2M_1\right) \dt\\
&\quad {} -2\int^x_0 e^{-\int^t_0 6\lambda_t^{2(0)} \dt'}\left(\frac{16}{9}M_3 - 3M_2 - \frac{13}{99}M_1\right)
\left(\frac{16}{3}g_3^2M_3 + 3g_2^2M_2 + \frac{13}{15}g_1^2M_1\right) \dt \label{technic2}
\end{align}
Therefore,
\allowdisplaybreaks
\begin{align}
&\left(m^2_{L_3} + m^2_{H_u}\right)^{(0)}\nonumber \\
&\qquad = m^2_{L_3}(0) + m^2_{H_u}(0) + \frac{1}{2}D_t(0)\left(e^{\int^x_0 6\lambda_t^{2(0)}\dt}-1\right)
- \int^x_0 \left(\frac{16}{3}g_3^2M_3^2 -3 g_2^2M_2^2- \frac{1}{3}g_1^2M_1^2\right) \dt \nonumber \\
&\qquad \quad {} -\frac{1}{2}A_t^2(0)e^{\int^x_0 6\lambda_t^{2(0)} \dt} - A_t(0)e^{\int^x_0 6\lambda_t^{2(0)}\dt} \int^x_0 \left(\frac{16}{3}g_3^2M_3 + 3g_2^2M_2 + \frac{13}{15}g_1^2M_1 \right)\dt \nonumber \\
&\qquad \quad {} -e^{\int^x_0 6\lambda_t^{2(0)}\dt} \int^x_0 e^{-\int^t_0 6\lambda_t^{2(0)}\dt} \left(
\frac{16}{3}g_3^2M_3^2 + 3g_2^2M_2^2 + \frac{13}{15}g_1^2M_1^2 \right)\dt \nonumber \\
&\qquad \quad {} -e^{\int^x_0 6\lambda_t^{2(0)}\dt} \int^x_0 e^{-\int^t_0 6\lambda_t^{2(0)}\dt}
\left(\frac{16}{9}M_3 - 3M_2 -\frac{13}{99}M_1\right)\left(\frac{16}{3}g_3^2M_3 + 3g_2^2M_2 + \frac{13}{15}g_1^2M_1\right)\dt \nonumber \\
&\qquad \quad {} +e^{\int^x_0 6\lambda_t^{2(0)}\dt} \left(\frac{16}{9}M_3-3M_2-\frac{13}{99}M_1\right)\int^x_0 e^{-\int^t_0 6\lambda_t^{2(0)} \dt'}
\left(\frac{16}{3}g_3^2M_3 + 3g_2^2M_2 + \frac{13}{15}g_1^2M_1 \right) \dt \nonumber\\
&\qquad \quad {} +\frac{1}{2} e^{2\int^x_0 6\lambda_t^{2(0)} \dt}
\left[A_t(0) + \int^x_0 e^{-\int^t_0 6\lambda_t^{2(0)} \dt'}
\left(\frac{13}{3}g_3^2M_3 + 3g_2^2M_2 + \frac{13}{15}g_2^2M_1\right) \dt 
\right]^2 \label{mlhu0} \\
&\qquad = m_{L_3}^2(0) + m_{H_u}^2(0) + \sum_X \alpha_X^{(0)}\fn{X}{0} \text{ with } X \in \{D_t,A_t^2,M_i,M_iM_j\,|\,i,j=1,2,3 \}
\end{align}
where the coefficients $\alpha^{(0)}_X$ are functions of $\fn{g_1}{0},\fn{g_2}{0},\fn{g_3}{0},\fn{\lambda_t}{0}$ and $x$. For example,
\begin{equation}
\alpha_{A_tM_3}^{(0)} = \frac{16}{3}e^{\int^x_0 6\lambda_t^{2(0)}\,dt}\int^x_0 \frac{\fn{g_3^2}{0}}{1+3\fn{g_3^2}{0}t}\,dt +\frac{13}{3}e^{2\int^x_0 6\lambda_t^{2(0)}\,dt}\int^x_0 e^{-\int^t_0 6\lambda_t^{2(0)}\,dt'}\frac{\fn{g_3^2}{0}}{1+3\fn{g_3^2}{0}t}\,dt
\end{equation}
where $\lambda_t$ is given by Eq.\,\eqref{Sollt}. We evaluated these integrals numerically, and the resulting numerical values of $\alpha_i^{(0)}$ for $\tan\beta = 5, 10, 15, 20, 30$ and AD scale $= 10^8, 10^9, 10^{10} \gev$ are provided in Tables \ref{tanb5} to \ref{tanb30}.

\subsection{First order solution}\label{a2-2}
Substituting Eqs.\,\eqref{RGEmlt} to \eqref{RGEmltau}, Eq.\,\eqref{RGEmlhu} can be written as 
	\begin{equation}
	\frac{d}{dx}\left(m_{L_3}^{2}+m_{H_u}^{2}\right)^{(1)} = \frac{33}{61}\frac{d}{dx}D_t^{(1)}
	- \frac{15}{61}\frac{d}{dx}D_b + \frac{19}{61}\frac{d}{dx}D_\tau +\frac{192}{61} g_3^2M_3^2 +\frac{144}{61} g_2^2M_2^2 + \frac{192}{305} g_1^2M_1^2 \label{firstorder}
	\end{equation}
	and solved as
	\begin{align}
\nonumber \left[m_{L_3}^{2}(x)+m_{H_u}^{2}(x)\right]^{(1)} &= m_{L_3}^2(0) + m_{H_u}^2(0)\\
\nonumber &\quad +  \frac{33}{61}\left[D_t^{(1)}(x)-D_t(0)\right] - \frac{15}{61}\left[D_b^{(1)}(x)-D_b(0)\right] + \frac{19}{61}\left[D_\tau^{(1)}(x)-D_\tau(0)\right]\\
& \quad +\int^x_0 \left(\frac{192}{61} g_3^2M_3^2  -\frac{144}{61} g_2^2M_2^2  +\frac{192}{305} g_1^2M_1^2 \right) \, dt \label{intermediate-mlhu}
\end{align}
Solutions of Eqs.\,\eqref{RGElt} to \eqref{RGEmltau} in the form of Eq.\,\eqref{formula2} are	
	\begin{align}
	\frac{1}{\lambda_b^2} & = e^{\int^x_0 \left(\frac{16}{3}g_3^2 + 3g_2^2+\frac{7}{15}g_1^2 - \lambda_t^2\right) \dt} \left[\frac{1}{\lambda_b^2(0)} -6\int^x_0 e^{-\int^t_0 \left(\frac{16}{3}g_3^2 + 3g_2^2+\frac{7}{15}g_1^2 - \lambda_t^2\right) \dt'} \dt \right]\\
	A_b & = e^{\int^x_0 6\lambda_b^2\dt}\left[A_b(0) + \int^x_0 e^{-\int^t_0 6\lambda_b^2\dt'}\left(\lambda_t^2A_t + \frac{16}{3}g_3^2M_3 + 3g_2^2M_2 + \frac{7}{15}g_1^2M_1\right)\dt\right] \label{SolAb1}\\
	D_b & = e^{\int^x_0 6\lambda_b^2 \dt}\left[D_b(0)
	+ \int^x_0 e^{-\int^t_0 6\lambda_b^2 \dt'}
	\left(\lambda_t^2A_t^2+ \lambda_t^2D_t + 6\lambda_b^2A_b^2
	-\frac{32}{3}g_3^2M_3^2 -6g_2^2M_2^2 - \frac{14}{15}g_1^2M_1^2 \right) \dt \right]\hspace{.6cm}  \label{SolDb1} \\
	\frac{1}{\lambda_\tau^2} & = e^{\int^x_0  \left(3g_2^2+\frac{9}{5}g_1^2 - 3\lambda_b^2 \right)\dt} \left[\frac{1}{\lambda_\tau^2(0)} -4\int^x_0 e^{\int^t_0 \left(3g_2^2+\frac{9}{5}g_1^2 - 3\lambda_b^2\right) \dt'} \dt \right]\\
	A_\tau & = e^{\int^x_0 4\lambda_\tau^2\dt}\left[A_\tau(0) + \int^x_0 e^{-\int^t_0 4\lambda_\tau^2\dt'}\left(3\lambda_b^2A_b  - 3g_2^2M_2 - \frac{3}{5}g_1^2M_1\right)\dt\right] \label{SolAtau1}\\
	D_\tau & = e^{\int^x_0 4\lambda_\tau^2 \dt}\left[D_\tau(0) 
	+ \int^x_0 e^{-\int^t_0 4\lambda_\tau^2 \dt'}
	\left(3\lambda_b^2A_b^2+ 3\lambda_t^2D_b + 4\lambda_\tau^2A_\tau^2
	- 6g_2^2M_2^2 - \frac{18}{5}g_1^2M_1^2 \right) \dt \right] \label{SolDtau1}\\
	\frac{1}{\lambda_t^{2(1)}} & = e^{\int^x_0 \left(\frac{16}{3}g_3^2 + 3g_2^2+\frac{13}{15}g_1^2 - \lambda_b^2\right) \dt} \left[\frac{1}{\lambda_t^2(0)} -6\int^x_0 e^{\int^t_0 \left(\frac{16}{3}g_3^2 + 3g_2^2+\frac{13}{15}g_1^2 - \lambda_b^2\right) \dt'} \dt \right]\\
	A_t^{(1)} & = e^{\int^x_0 6\lambda_t^{2(1)}\dt}\left[A_t(0) + \int^x_0 e^{-\int^t_0 6\lambda_t^{2(1)}\dt'}\left(\lambda_b^2A_b + \frac{16}{3}g_3^2M_3 + 3g_2^2M_2 + \frac{13}{15}g_1^2M_1\right)\dt\right] \label{SolAt1}\\
	\nonumber D_t^{(1)} & = e^{\int^x_0 6\lambda_t^{2(1)} \dt}D_t(0)\\ 
	& \quad + e^{\int^x_0 6\lambda_t^{2(1)} \dt}\int^x_0 e^{-\int^t_0 6\lambda_t^{2(1)} \dt'}
	\left[\lambda_b^2\left(A_b^2+D_b\right) + 6\lambda_t^{2(1)}A_t^{2(1)}
	-\frac{32}{3}g_3^2M_3^2 -6g_2^2M_2^2 - \frac{26}{15}g_1^2M_1^2 \right] \dt \label{SolDt1}
\end{align}
Substituting Eqs.\,\eqref{SolAb1},\eqref{SolAtau1} and \eqref{SolAt1} into Eqs.\,\eqref{SolDb1}, \eqref{SolDtau1} and \eqref{SolDt1} respectively using the similar technique we used in Eqs.\,\eqref{technic1} and \eqref{technic2}, one can express Eq.\,\eqref{intermediate-mlhu} in an integral form with a few integrals. This can be expressed as
 \begin{gather}
\left[m_{L_3}^{2}(x)+m_{H_u}^{2}(x)\right]^{(1)} = m_{L_3}^2(0) + m_{H_u}^2(0) + \sum_X \alpha_X^{(1)}\fn{X}{0}  \label{expansion1}\\
\text{with } X \in \{D_\alpha, A_\alpha A_\beta, A_\alpha M_i, M_iM_j \,|\, \alpha,\beta = t,b,\tau \text{ and } i,j=1,2,3\} \nonumber
\end{gather} 
where the $\alpha_X^{(1)}$ are functions of $\fn{g_1}{0}, \fn{g_2}{0}, \fn{g_3}{0}, \fn{\lambda_t}{0}, \fn{\lambda_b}{0}, \fn{\lambda_\tau}{0}$ and $x$. After numerically evaluating the integrals in Eqs.\,\eqref{SolAb1}, \eqref{SolAtau1} and \eqref{SolAt1} and substituting the results into $D_b, D_\tau$ and $D_t^{(1)}$ in Eq.\,\eqref{expansion1}, one can obtain the numerical values of $\alpha_i^{(1)}$. The results are provided in Tables \ref{tanb5} to \ref{tanb30}.

\subsection{Numerical coefficients of the baryogenesis condition} \label{a2-3}
The numerical coefficients in Eqs.\,\eqref{expansion0} and \eqref{expansion1} are given in the following tables.


\begin{table}[htbp]
	\centering
	\setlength\extrarowheight{5pt}
	\begin{tabular}{c||c|c|c|c|c|c|c|c|c|c|c}
		\hline
		\multicolumn{12}{c}{$\tan\beta=5$}\\
		\hline
		AD scale  & $\alpha_{D_t}^{(0)}$ & $\alpha_{A_t^2}^{(0)}$ & $\alpha_{A_tM_1}^{(0)}$ & $\alpha_{A_tM_2}^{(0)}$ & $\alpha_{A_tM_3}^{(0)}$ & $\alpha_{M_1^2}^{(0)}$ & $\alpha_{M_2^2}^{(0)}$ & $\alpha_{M_3^2}^{(0)}$ & $\alpha_{M_1M_2}^{(0)}$ &$\alpha_{M_2M_3}^{(0)}$ & $\alpha_{M_3M_1}^{(0)}$ \\
		\hline
		$10^8$ GeV & 0.26 & 0.40 & 0.01 & 0.05 & 0.18 &  -0.05 & -0.35 & -0.08 & 0.00 & 0.01 & 0.00 \\
		$10^9$ GeV & 0.33 &  0.54 & 0.02 & 0.09 & 0.28 & -0.07 &-0.45 & -0.11 & 0.00 & 0.03 &0.01 \\
		$10^{10}$ GeV & 0.40 & 0.71 & 0.02 & 0.13 & 0.41 & -0.09 & -0.55 & -0.13 & 0.00 & 0.05 & 0.01  \\
		\hline
		\hline
		AD scale  & $\alpha_{D_t}^{(1)}$ & $\alpha_{A_t^2}^{(1)}$ & $\alpha_{A_tM_1}^{(1)}$ & $\alpha_{A_tM_2}^{(1)}$ & $\alpha_{A_tM_3}^{(1)}$ & $\alpha_{M_1^2}^{(1)}$ & $\alpha_{M_2^2}^{(1)}$ & $\alpha_{M_3^2}^{(1)}$ & $\alpha_{M_1M_2}^{(1)}$ &$\alpha_{M_2M_3}^{(1)}$ & $\alpha_{M_3M_1}^{(1)}$ \\
		\hline
		$10^8$ GeV & 0.26 & 0.40 & 0.01 & 0.05 & 0.18 & -0.05 & -0.33 & -0.08 & 0.01 & 0.02 & 0.00 \\
		$10^9$ GeV & 0.33 & 0.54 & 0.02 & 0.09 & 0.28  & -0.07 &-0.42& -0.11 &0.02 & 0.03 &0.01 \\
		$10^{10}$ GeV & 0.40 & 0.71 & 0.02 & 0.13 & 0.42 & -0.09 & -0.52 & -0.13 & 0.03 & 0.05 & 0.01 \\
		\hline		
	\end{tabular}
	\caption{Numerical coefficients of $m_{L_3}^{2} + m_{H_u}^{2}$ at zeroth and the first order for $\tan\beta=5$. FlexibleSUSY was used to get the low scale values of the couplings $\fn{\lambda_t}{0}=0.81,\fn{\lambda_b}{0}=0.06,\fn{\lambda_\tau}{0}=0.05,\fn{g_1}{0}=0.47,\fn{g_2}{0}=0.63,\fn{g_3}{0}=0.97$ at $m_s = 6 \tev $.}
	\label{tanb5}
\end{table}

\begin{table}[htbp]
	\centering
	\setlength\extrarowheight{5pt}
	\begin{tabular}{c||c|c|c|c|c|c|c|c|c|c|c}
		\hline
		\multicolumn{12}{c}{$\tan\beta=10$}\\
		\hline
		AD scale  & $\alpha_{D_t}^{(0)}$ & $\alpha_{A_t^2}^{(0)}$ & $\alpha_{A_tM_1}^{(0)}$ & $\alpha_{A_tM_2}^{(0)}$ & $\alpha_{A_tM_3}^{(0)}$ & $\alpha_{M_1^2}^{(0)}$ & $\alpha_{M_2^2}^{(0)}$ & $\alpha_{M_3^2}^{(0)}$ & $\alpha_{M_1M_2}^{(0)}$ &$\alpha_{M_2M_3}^{(0)}$ & $\alpha_{M_3M_1}^{(0)}$ \\
		\hline
		$10^8$ GeV & 0.38 & 0.01 & 0.05 & 0.17 & 0.25 & -0.05 & -0.35 & -0.08 & 0.00 & 0.01 & 0.00 \\
		$10^9$ GeV & 0.51 & 0.01 & 0.08 & 0.27 & 0.31 & -0.07 &-0.45 & -0.11 & 0.00 & 0.03 & 0.00 \\
		$10^{10}$ GeV & 0.66 & 0.02 & 0.13 & 0.39 & 0.38 & -0.09 & -0.55 & -0.13 & 0.00 & 0.05 & 0.01 \\
		\hline
		\hline
		AD scale  & $\alpha_{D_t}^{(1)}$ & $\alpha_{A_t^2}^{(1)}$ & $\alpha_{A_tM_1}^{(1)}$ & $\alpha_{A_tM_2}^{(1)}$ & $\alpha_{A_tM_3}^{(1)}$ & $\alpha_{M_1^2}^{(1)}$ & $\alpha_{M_2^2}^{(1)}$ & $\alpha_{M_3^2}^{(1)}$ & $\alpha_{M_1M_2}^{(1)}$ &$\alpha_{M_2M_3}^{(1)}$ & $\alpha_{M_3M_1}^{(1)}$ \\
		\hline
		$10^8$ GeV & 0.38 & 0.01 & 0.05 & 0.17 & 0.25 &  -0.05 & -0.33 & -0.08 & 0.01 & 0.01 & 0.00  \\
		$10^9$ GeV & 0.51 & 0.01 & 0.08 & 0.27 & 0.32 & -0.07 & -0.42 & -0.11 & 0.02 & 0.03 & 0.01 \\
		$10^{10}$ GeV & 0.67 & 0.02 & 0.13 & 0.39 & 0.38 & -0.09 & -0.52 & -0.13 & 0.03 & 0.05 & 0.01 \\
		\hline		
	\end{tabular}
	\caption{Numerical coefficients of $m_{L_3}^{2} + m_{H_u}^2$ at zeroth and first order for $\tan\beta=10$. Used $\fn{\lambda_t}{0}=0.80,\fn{\lambda_b}{0}=0.12,\fn{\lambda_\tau}{0}=0.10,\fn{g_1}{0}=0.47,\fn{g_2}{0}=0.63,\fn{g_3}{0}=0.97$ at $m_s = 6 \tev $.}
	\label{tanb10}
\end{table}

\clearpage	
	
\begin{table}[htbp]
	\centering
	\setlength\extrarowheight{6pt}
	\begin{tabular}{c||c|c|c|c|c|c|c|c|c|c|c}
		\hline
		\multicolumn{12}{c}{$\tan\beta=15$}\\
		\hline
		AD scale  & $\alpha_{D_t}^{(0)}$ & $\alpha_{A_t^2}^{(0)}$ & $\alpha_{A_tM_1}^{(0)}$ & $\alpha_{A_tM_2}^{(0)}$ & $\alpha_{A_tM_3}^{(0)}$ & $\alpha_{M_1^2}^{(0)}$ & $\alpha_{M_2^2}^{(0)}$ & $\alpha_{M_3^2}^{(0)}$ & $\alpha_{M_1M_2}^{(0)}$ &$\alpha_{M_2M_3}^{(0)}$ & $\alpha_{M_3M_1}^{(0)}$ \\
		\hline
		$10^8$ GeV & 0.36 & 0.01 & 0.05 & 0.16 & 0.24 & -0.05 & -0.35 & -0.08 & 0.00 & 0.01 & 0.00 \\
		$10^9$ GeV & 0.48 & 0.01 & 0.08 & 0.25 & 0.30 & -0.07 & -0.44 & -0.10 & 0.00 & 0.03 & 0.00 \\
		$10^{10}$ GeV & 0.62 & 0.02 & 0.12 & 0.37 & 0.36 & -0.09 & -0.55 & -0.13 & 0.00 & 0.04 & 0.01  \\
		\hline
		AD scale  & $\alpha_{D_t}^{(1)}$ & $\alpha_{A_t^2}^{(1)}$ & $\alpha_{A_tM_1}^{(1)}$ & $\alpha_{A_tM_2}^{(1)}$ & $\alpha_{A_tM_3}^{(1)}$ & $\alpha_{M_1^2}^{(1)}$ & $\alpha_{M_2^2}^{(1)}$ & $\alpha_{M_3^2}^{(1)}$ & $\alpha_{M_1M_2}^{(1)}$ &$\alpha_{M_2M_3}^{(1)}$ & $\alpha_{M_3M_1}^{(1)}$ \\
		\hline
		$10^8$ GeV & 0.36 & 0.01 & 0.05 & 0.16 & 0.24 & -0.05 & -0.33 & -0.08 & 0.01 & 0.01 & 0.00 \\
		$10^9$ GeV & 0.49 & 0.01 & 0.08 & 0.25 & 0.30 & -0.07 &-0.42& -0.10 & 0.02 & 0.03 & 0.00 \\
		$10^{10}$ GeV & 0.63 & 0.02 & 0.12 & 0.37 & 0.36 & -0.09 & -0.51 & -0.13 & 0.03 & 0.04 & 0.01 \\
		\hline
	\end{tabular}
	\caption{Numerical coefficients of $m_{L_3}^2 + m_{H_u}^{2}$ at zeroth and first order for $\tan\beta=15$. Used $\fn{\lambda_t}{0}=0.79,\fn{\lambda_b}{0}=0.18,\fn{\lambda_\tau}{0}=0.15,\fn{g_1}{0}=0.47,\fn{g_2}{0}=0.63,\fn{g_3}{0}=0.97$ at $m_s = 6 \tev $.}
	\label{tanb15}
\end{table}	

\begin{table}[htbp]
	\centering
	\setlength\extrarowheight{5pt}
	\begin{tabular}{c||c|c|c|c|c|c|c|c|c|c|c}
		\hline
		\multicolumn{12}{c}{$\tan\beta=20$}\\
		\hline
		AD scale  & $\alpha_{D_t}^{(0)}$ & $\alpha_{A_t^2}^{(0)}$ & $\alpha_{A_tM_1}^{(0)}$ & $\alpha_{A_tM_2}^{(0)}$ & $\alpha_{A_tM_3}^{(0)}$ & $\alpha_{M_1^2}^{(0)}$ & $\alpha_{M_2^2}^{(0)}$ & $\alpha_{M_3^2}^{(0)}$ & $\alpha_{M_1M_2}^{(0)}$ &$\alpha_{M_2M_3}^{(0)}$ & $\alpha_{M_3M_1}^{(0)}$ \\
		\hline
		$10^8$ GeV & 0.36 & 0.01 & 0.05 & 0.16 & 0.24 & -0.05 & -0.35 & -0.08 & 0.00 & 0.01 & 0.00 \\
		$10^9$ GeV & 0.49 & 0.01 & 0.08 & 0.25 & 0.30 & -0.07 & -0.44 & -0.10 & 0.00 & 0.03 & 0.00 \\
		$10^{10}$ GeV & 0.63 & 0.02 & 0.12 & 0.37 & 0.37 & -0.09 & -0.55 & -0.13 & 0.00 & 0.04 & 0.01 \\
		\hline
		\hline
		AD scale  & $\alpha_{D_t}^{(1)}$ & $\alpha_{A_t^2}^{(1)}$ & $\alpha_{A_tM_1}^{(1)}$ & $\alpha_{A_tM_2}^{(1)}$ & $\alpha_{A_tM_3}^{(1)}$ & $\alpha_{M_1^2}^{(1)}$ & $\alpha_{M_2^2}^{(1)}$ & $\alpha_{M_3^2}^{(1)}$ & $\alpha_{M_1M_2}^{(1)}$ &$\alpha_{M_2M_3}^{(1)}$ & $\alpha_{M_3M_1}^{(1)}$ \\
		\hline
		$10^8$ GeV & 0.37 & 0.01 & 0.05 & 0.16 & 0.25 & -0.05 & -0.33 & -0.08 & 0.01 & 0.01 & 0.00 \\
		$10^9$ GeV & 0.49 & 0.01 & 0.08 & 0.26 & 0.31 & -0.07 & -0.42 & -0.10 & 0.02 & 0.03 & 0.01 \\
		$10^{10}$ GeV & 0.64 & 0.02 & 0.12 & 0.37 & 0.37 & -0.09 & -0.51 & -0.13 & 0.03 & 0.05 & 0.01 \\
		\hline		
		& $\alpha_{A_bM_2}^{(1)}$ &  $\alpha_{A_\tau^2}^{(1)}$ & $\alpha_{D_\tau}^{(1)}$ & $\alpha_{A_\tau M_2}^{(1)}$  & & & & & & &  \\
		\hline
		$10^8$ GeV & 0.00 & 0.01 & 0.01 & 0.00 & & & & & & & \\
		$10^9$ GeV & 0.01 & 0.01 & 0.01 & 0.00 & & & & & & & \\
		$10^{10}$ GeV & 0.01 & 0.01 & 0.01 & 0.01 & & & & & & & \\ 
		\hline
	\end{tabular}
	\caption{Numerical coefficients of $m_{L_3}^2 + m_{H_u}^{2}$ at zeroth and first order for $\tan\beta=20$. Used $\fn{\lambda_t}{0}=0.79,\fn{\lambda_b}{0}=0.24,\fn{\lambda_\tau}{0}=0.20,\fn{g_1}{0}=0.47,\fn{g_2}{0}=0.63,\fn{g_3}{0}=0.96$ at $m_s = 6 \tev $.}
	\label{tanb20}
\end{table}	

\clearpage

\begin{table}[htbp]
	\centering
	\setlength\extrarowheight{5pt}
	\begin{tabular}{c||c|c|c|c|c|c|c|c|c|c|c}
		\hline
		\multicolumn{12}{c}{$\tan\beta=30$}\\
		\hline
		AD scale  & $\alpha_{D_t}^{(0)}$ & $\alpha_{A_t^2}^{(0)}$ & $\alpha_{A_tM_1}^{(0)}$ & $\alpha_{A_tM_2}^{(0)}$ & $\alpha_{A_tM_3}^{(0)}$ & $\alpha_{M_1^2}^{(0)}$ & $\alpha_{M_2^2}^{(0)}$ & $\alpha_{M_3^2}^{(0)}$ & $\alpha_{M_1M_2}^{(0)}$ &$\alpha_{M_2M_3}^{(0)}$ & $\alpha_{M_3M_1}^{(0)}$ \\
		\hline
		$10^8$ GeV & 0.36 & 0.01 & 0.05 & 0.16 & 0.24 & -0.05 & -0.35 & -0.08 & 0.00 & 0.01 & 0.00 \\
		$10^9$ GeV & 0.49 & 0.01 & 0.08 & 0.25 & 0.30 & -0.07 & -0.44 & -0.10 & 0.00 & 0.03 & 0.00 \\
		$10^{10}$ GeV & 0.63 & 0.02 & 0.12 & 0.37 & 0.37 & -0.09 & -0.55 & -0.13 & 0.00 & 0.04 & 0.01  \\
		\hline
		\hline
		AD scale  & $\alpha_{D_t}^{(1)}$ & $\alpha_{A_t^2}^{(1)}$ & $\alpha_{A_tM_1}^{(1)}$ & $\alpha_{A_tM_2}^{(1)}$ & $\alpha_{A_tM_3}^{(1)}$ & $\alpha_{M_1^2}^{(1)}$ & $\alpha_{M_2^2}^{(1)}$ & $\alpha_{M_3^2}^{(1)}$ & $\alpha_{M_1M_2}^{(1)}$ &$\alpha_{M_2M_3}^{(1)}$ & $\alpha_{M_3M_1}^{(1)}$ \\
		\hline
		$10^8$ GeV & 0.37 & 0.01 & 0.05 & 0.16 & 0.25 & -0.05 & -0.33 & -0.08 & 0.01 & 0.02 & 0.00 \\
		$10^9$ GeV & 0.50 & 0.01 & 0.08 & 0.26 & 0.31 & -0.07 & -0.42 & -0.10 & 0.02 & 0.03 & 0.01 \\
		$10^{10}$ GeV & 0.65 & 0.02 & 0.12 & 0.38 & 0.37 & -0.09 & -0.51 & -0.13 & 0.04 & 0.05 & 0.01 \\
		\hline		
		& $\alpha_{A_b^2}^{(1)}$ & $\alpha_{A_bA_t}^{(1)}$ & $\alpha_{A_bM_1}^{(1)}$ & $\alpha_{A_bM_2}^{(1)}$ & $\alpha_{A_bM_3}^{(1)}$ & $\alpha_{A_\tau^2}^{(1)}$ & $\alpha_{D_\tau}^{(1)}$ & $\alpha_{A_\tau M_1}^{(1)}$& $\alpha_{A_\tau M_2}^{(1)}$ & &   \\
		\hline
		$10^8$ GeV & 0.00 & 0.00 & 0.00 & 0.01 & 0.00 & 0.01 & 0.01 & 0.00 & 0.01 & &\\
		$10^9$ GeV & 0.00 & 0.01 & 0.00 & 0.01 & 0.00 & 0.02 & 0.02 & 0.00 & 0.01 & &\\
		$10^{10}$ GeV & 0.01 & 0.01 & 0.01 & 0.02 & 0.01 & 0.02 & 0.02 & 0.01 & 0.02 & &\\ 
		\hline
	\end{tabular}
	\caption{Numerical coefficients of $m_{L_3}^2 + m_{H_u}^{2}$ at zeroth and first order for $\tan\beta=30$. Used $\fn{\lambda_t}{0}=0.79,\fn{\lambda_b}{0}=0.35,\fn{\lambda_\tau}{0}=0.31,\fn{g_1}{0}=0.47,\fn{g_2}{0}=0.63,\fn{g_3}{0}=0.96$ at $m_s = 6 \tev $.}
	\label{tanb30}
\end{table}

\end{appendices}

\bibliographystyle{utphys}
\bibliography{bib/bib}

\end{document}